\documentclass[twocolumn]{aastex631}

\usepackage{color}
\usepackage{xspace}

\newcommand{\lext}{{\em LEIA}\xspace}

\newcommand{\NAOC}{National Astronomical Observatories, Chinese Academy of Sciences, Beijing, 100101, P.R. China}
\newcommand{\NAOD}{National Astronomical Data Center of China, Beijing, 100101, P.R. China}
\newcommand{\UCAS}{School of Astronomy and Space Sciences, University of Chinese Academy of Sciences, Beijing, 100049, P.R. China}
\newcommand{\IHEP}{Key Laboratory of Particle Astrophysics, Institute of High Energy Physics, Chinese Academy of Sciences, Beijing, 100049, P.R. China}
\newcommand{\NSSC}{National Space Science Center, Chinese Academy of Sciences, Beijing, 100190, P.R. China}
\newcommand{\SITP}{Shanghai Institute of Technical Physics, Chinese Academy of Sciences, Shanghai, 200083, P.R. China}
\newcommand{\STL}{Science and Technology on Low-Light-Level Night Vision Laboratory, Xi'an, 710065, P.R. China}
\newcommand{\microSat}{Innovation Academy for Microsatellites, Chinese Academy of Sciences, Shanghai, 201210, P.R. China}
\newcommand{\NNVT}{North Night Vision Technology Co., LTD, Nanjing, P.R. China}
\newcommand{\LU}{School of Physics and Astronomy, University of Leicester, University Road, Leicester LE1 7RH, UK}
\newcommand{\MPE}{Max-Planck-Institut für extraterrestrische Physik, Gießenbachstraße 1, 85748 Garching, Germany}
\newcommand{\ESA}{European Space Agency (ESA), European Space Research and Technology Centre (ESTEC), Keplerlaan 1, 2201 AZ Noordwijk, The Netherlands}

\def\ergs{${\rm erg\,s^{-1}\,cm^{-2}}$\xspace}

\begin{document}

\title{First wide field-of-view X-ray observations by a lobster eye focusing telescope in orbit}

\correspondingauthor{Z.X. Ling, S.L. Sun, W. Yuan}
\email{lingzhixing@nao.cas.cn, palm\_sun@mail.sitp.ac.cn, wmy@nao.cas.cn}

%\author[0000-0002-0786-7307]{Chen Zhang}
\author{C. Zhang}
\affiliation{\NAOC}
\affiliation{\UCAS}
\author{Z.X. Ling}
\affiliation{\NAOC}
\affiliation{\UCAS}
\author{X.J. Sun}
\affiliation{\SITP}
\author{S.L. Sun}
\affiliation{\SITP}
%\author{W. Yuan}
%\affiliation{NAOC, CAS}
%\affiliation{UCAS}
\author{Y. Liu}
\affiliation{\NAOC}
\author{Z.D. Li}
\affiliation{\SITP}
\author{Y.L. Xue}
\affiliation{\SITP}
\author{Y.F. Chen}
\affiliation{\SITP}
%\author{Y. Liu}
%\affiliation{NAOC, CAS}
\author{Y.F. Dai}
\affiliation{\NAOC}
\author{Z.Q. Jia}
\affiliation{\NAOC}
\author{H.Y. Liu}
\affiliation{\NAOC}
\author{X.F. Zhang}
\affiliation{\microSat}
\author{Y.H. Zhang}
\affiliation{\microSat}
\author{S.N. Zhang}
\affiliation{\IHEP}
\affiliation{\NAOC}
\affiliation{\UCAS}
%\author{W. Yuan}
%\affiliation{NAOC, CAS}
%\affiliation{UCAS}
%\altaffiliation{AASTeX v6+ programmer}

%SITP
\author{F.S. Chen}
\affiliation{\SITP}
\author{Z.W. Cheng}
\affiliation{\SITP}
\author{W. Fu}
\affiliation{\SITP}
\author{Y.X. Han}
\affiliation{\SITP}
\author{H. Li}
\affiliation{\SITP}
\author{J.F. Li}
\affiliation{\SITP}
\author{Y. Li}
\affiliation{\SITP}
\author{P.R. Liu}
\affiliation{\SITP}
\author{X.H. Ma}
\affiliation{\SITP}
\author{Y.J. Tang}
\affiliation{\SITP}
\author{C.B. Wang}
\affiliation{\SITP}
\author{R.J. Xie}
\affiliation{\SITP}
\author{A.L. Yan}
\affiliation{\SITP}
\author{Q. Zhang}
\affiliation{\SITP}

%NNVT
\author{B.W. Jiang}
\affiliation{\NNVT}
\author{G. Jin}
\affiliation{\NNVT}
\author{L.H. Li}
\affiliation{\NNVT}
\author{X.B. Qiu}
\affiliation{\NNVT}
\author{D.T. Su}
\affiliation{\NNVT}
\author{J.N. Sun}
\affiliation{\NNVT}
\author{Z. Xu}
\affiliation{\NNVT}
\author{S.K. Zhang}
\affiliation{\NNVT}
\author{Z. Zhang}
\affiliation{\NNVT}
\author{N. Zhang}
\affiliation{\STL}
%microsat
\author{X.Z. Bi}
\affiliation{\microSat}
\author{Z.M. Cai}
\affiliation{\microSat}
\author{J.W. He}
\affiliation{\microSat}
\author{H.Q. Liu}
\affiliation{\microSat}
\author{X.C. Zhu}
\affiliation{\microSat}

%NAOC
\author{H.Q. Cheng}
\affiliation{\NAOC}
\author{C.Z. Cui}
\affiliation{\NAOC}
\affiliation{\NAOD}
\author{D.W. Fan}
\affiliation{\NAOC}
\affiliation{\NAOD}
\author{H.B. Hu}
\affiliation{\NAOC}
\author{M.H. Huang}
\affiliation{\NAOC}
\author{C.C. Jin}
\affiliation{\NAOC}
\affiliation{\UCAS}
\author{D.Y. Li}
\affiliation{\NAOC}
\author{H.W. Pan}
\affiliation{\NAOC}
\author{W.X. Wang}
\affiliation{\NAOC}
\author{Y.F. Xu}
\affiliation{\NAOC}
\affiliation{\NAOD}
\author{X. Yang}
\affiliation{\NAOC}
\author{B. Zhang}
\affiliation{\NAOC}
\author{M. Zhang}
\affiliation{\NAOC}
\author{W.D. Zhang}
\affiliation{\NAOC}
\author{D.H. Zhao}
\affiliation{\NAOC}

%nssc
\author{M. Bai}
\affiliation{\NSSC}
\author{Z. Ji}
\affiliation{\NSSC}
\author{Y.R. Liu }
\affiliation{\NSSC}
\author{F.L. Ma}
\affiliation{\NSSC}
\author{J. Su}
\affiliation{\NSSC}
\author{J.Z. Tong}
\affiliation{\NSSC}

%ihep
\author{Y.S. Wang}
\affiliation{\IHEP}
\author{Z.J. Zhao}
\affiliation{\IHEP}

%LU
\author{C. Feldman}
\affiliation{\LU}
\author{P. O'Brien}
\affiliation{\LU}
\author{J.P. Osborne}
\affiliation{\LU}
\author{R. Willingale}
\affiliation{\LU}

%MPE
\author{V. Burwitz}
\affiliation{\MPE}
\author{G. Hartner}
\affiliation{\MPE}
\author{A. Langmeier}
\affiliation{\MPE}
\author{T. M\"{u}ller}
\affiliation{\MPE}
\author{S. Rukdee}
\affiliation{\MPE}
\author{T. Schmidt}
\affiliation{\MPE}

%esa
\author{E. Kuulkers}
\affiliation{\ESA}
%CAS
\author{W. Yuan}
\affiliation{\NAOC}
\affiliation{\UCAS}

%\author{more authors}
%\affiliation{SITP}
%\author{more authors}
%\affiliation{NNVT}
%\author{more authors}
%\affiliation{MicroSat}
%\author{more authors}
%\affiliation{NAOC}
%\author{more authors}
%\affiliation{EPMC}
%\author{more authors}
%\affiliation{IHEP}
%\author{more authors}
%\affiliation{LU}
%\author{more authors}
%\affiliation{MPE}
%\author{E. Kuulkers}
%\affiliation{ESA}
%\author{S.L. Sun}
%\affiliation{SITP, CAS}
%\author{W. Yuan}
%\affiliation{NAOC, CAS}
%\affiliation{UCAS}
%

%\affiliation{AAS Journals Associate Editor-in-Chief}



\begin{abstract}
As a novel X-ray focusing technology, lobster eye micro-pore optics (MPO) feature both a wide observing field of view and true imaging capability, promising sky monitoring with significantly improved sensitivity and spatial resolution in soft X-rays.
Since first proposed by \citet{1979ApJ...233..364A}, the optics have been extensively studied, developed and trialed over the past decades. 
In this Letter, we report on the first-light results from a flight experiment of the {\em Lobster Eye Imager for Astronomy} (\lext), a pathfinder of the wide-field X-ray telescope of the Einstein Probe mission. 
The piggyback imager, launched in July 2022, has a mostly un-vignetted field of view of $18.6^\circ \times 18.6^\circ $. 
Its spatial resolution is in the range of 4--7\,arcmin in FWHM and the focal spot effective area is 2--3\,cm$^2$, both showing only mild fluctuations across the field of view.  
We present images of the Galactic center region, Sco X-1 and the diffuse Cygnus Loop nebular taken in snapshot observations over 0.5--4\,keV. 
These are truly wide-field X-ray images of celestial bodies observed, for the first time, by a focusing imaging telescope.
Initial analyses of the in-flight data show excellent agreement between the observed images and the on-ground calibration and simulations.
The instrument and its characterization are briefly described, as well as the flight experiment. 
The results provide a solid basis for the development of the present and proposed wide-field X-ray missions using lobster eye MPO.
\end{abstract}


%\keywords{X-ray telescope (251) --- (1736) --- History of astronomy(1868) --- Interdisciplinary astronomy(804)}


\section{Introduction} 
\label{sec:intro}
Wide-field monitoring of the X-ray sky plays an indispensable role in understanding the dynamic X-ray universe. 
The detection and alert of transients and observation of variable sources in large numbers require large field of view (FoV) X-ray detectors, preferably an All-Sky Monitor (ASM).
Tremendous advances in this field have been made over the past several decades by a series of X-ray ASMs.
Examples of these range from previous instruments Vela 5, Ariel V, Ginga-ASM, RXTE-ASM, HETE2 and Beppo-SAX-WFC, right up to the ones currently in operation, including Swift-BAT and MAXI. 
%\footnote{A few high-energy missions with a moderate FoV also have X-ray monitoring capability to some extent, such as INTEGRAL/IBIS and {\it Insight-HXMT} \citep{hxmt2020SCPMA..6349502Z}, while others with a small FoV can survey the sky by scanning at a low cadence, such as the XMM-Newton slew survey and eROSITA.}.  
With a 1.4\,sr FoV, Swift/BAT \citep{2004ApJ...611.1005G} detects GRBs and other bright fast transients in the 15-150 keV band, whilst MAXI \citep{MAXI2009PASJ...61..999M} monitors bright X-ray sources and transients primarily in our Galaxy in 2--30 keV via scanning almost the whole sky every 92 minutes.

In recent years some interesting extragalactic transients have been discovered and call for characterisation in large numbers, such as Gamma-Ray Bursts (GRB) beyond redshift 6, supernova shock breakouts, and tidal disruption events (TDE). 
The majority of these populations are at least 1--2 orders of magnitude fainter than the sensitivity of the current ASMs in orbit, however \citep{2015JHEAp...7....2G}. 
Specifically, to detect them a sensitivity level of several milliCrab (mCrab) for 1000\,s exposure would be desirable. Moreover, the emission of some of these new transients peaks in the soft X-ray band below a few keV, in contrast to the mostly hard X-ray bandpass of the existing ASMs. Most remarkably, the detection of the GRB\,170817 by Fermi and INTEGRAL as the X/$\gamma$-ray counterpart of the gravitational wave event GW\,170817 \citep{2017ApJ...848L..13A}
%, resulting from a binary neutron star merger, 
highlights further the great scientific potential of wide-field X-ray sky monitoring beyond the current horizon, particularly in the era of multi-messenger time-domain astronomy.  
    
So far all the X-ray ASMs are based on non-focusing techniques, e.g. pinhole/slit camera, collimator, or coded mask \citep{1987SSRv...45..269H}.
Their detecting sensitivity is rather limited, mostly caused by the high background level, resulting from their large point-spread function (PSF) profile, i.e. low angular resolution, as well as the multiplex nature of the imaging capability \citep{2017SSRv..207...63W}.
A preferable way to improve both sensitivity and spatial resolution is to use X-ray focusing optics \citep{2009xrda.book.....F}.

\citet{1979ApJ...233..364A} first proposed a design of X-ray ASM based on the imaging optics of the reflective eyes of lobsters.
This design enables grazing incidence reflection, and hence focusing, of soft X-rays by the smooth, reflective walls of many tiny square pores, which are densely packed across a sphere and all pointed toward a common center of curvature.
This is the so-called Micro-Pore Optics (MPO) \citep[see e.g.][for a recent review]{2022arXiv220807149H}.
The FoV of a Lobster Eye (LE) optic, which is the solid angle subtended by the optic plate to the curvature center, is limited only by the optic size for a given curvature radius. 
Since the MPO is spherically symmetric in essentially all directions, theoretically,
an idealized LE optic is almost free from vignetting except near the edge of the FoV. 
Over the past decades the LE optics has been studied extensively by several groups \citep[e.g.][]{1989RScI...60.1026W,1991RScI...62.1542C,1992SPIE.1546...41F,1992ApOpt..31.7339K,1993NIMPA.324..404F,1996ApOpt..35.4420P,1999NIMPA.431..356B,2014SPIE.9144E..4EZ,2015RScI...86g1301C,2016SPIE.9905E..1YW,2017SPIE10567E..19H}, and conceptual designs for LE X-ray ASMs have also been proposed \citep[e.g.][]{1996MNRAS.279..733P,2002SPIE.4497..115F}. 
In practice, a number of realistic missions with LE ASM have been proposed \citep[e.g.][]{2016grbu.book..237Y,2020SPIE11444E..2LO}.

The first instrument built for a formal mission is MIXS \citep{2020SSRv..216..126B} aboard BepiColombo, which consists of a $1^\circ$-FoV Wolter telescope and a $10^\circ$-FoV collimator, both built from MPO optic, and its first light is expected in several years time when the mission arrives at Mercury. Several X-ray telescopes based on LE MPO are under development and due to be launched in next few years, including SVOM-MXT \citep{2016SPIE.9905E..4LG,2022SPIE121811P..7RF} and SMILE-SXI \citep{2016AGUFMSM44A..04S}.

The Einstein Probe\footnote{Einstein Probe is a mission of the Chinese Academy of Sciences (CAS) in collaboration with the European Space Agency (ESA) and the Max-Planck-Institute for extraterrestrial Physics (MPE). 
} 
(EP) \citep{2018SPIE10699E..25Y,2022arXiv220909763Y}
is a time-domain astrophysics mission to discover cosmic high-energy transients and monitor variable objects.
It features a Wide-field X-ray Telescope (WXT) in 0.5--4\,keV consisting of 12 identical modules based on the LE MPO technique (Ling et al. in prep.) and a Follow-up X-ray Telescope (FXT) in 0.3--10\,keV \citep{2020SPIE11444E..5BC}.

To verify the in-orbit performance of the EP-WXT in advance and to optimize the instrumental parameters and conditions in operation, a complete test module of WXT was launched into orbit, as an EP-WXT pathfinder.
With a FoV of $18.6^\circ \times 18.6^\circ$, this instrument is a truly wide-field X-ray imager, termed {\em Lobster Eye Imager for Astronomy} (\lext).
Here we report on the initial results from the \lext experiment, which are the first truely wide-field X-ray images of celestial bodies ever taken by a focusing imaging X-ray telescope with one-shot exposures.
The instrument and the in-flight experiment are described in Section \ref{sec:instr} and \ref{sec:exp}, respectively, followed by their implications and conclusion in Section \ref{sec:con}.

\section{Description and characterization of the instrument} 
\label{sec:instr}

\subsection{Instrument description}

\begin{figure*}[ht!]
\plottwo{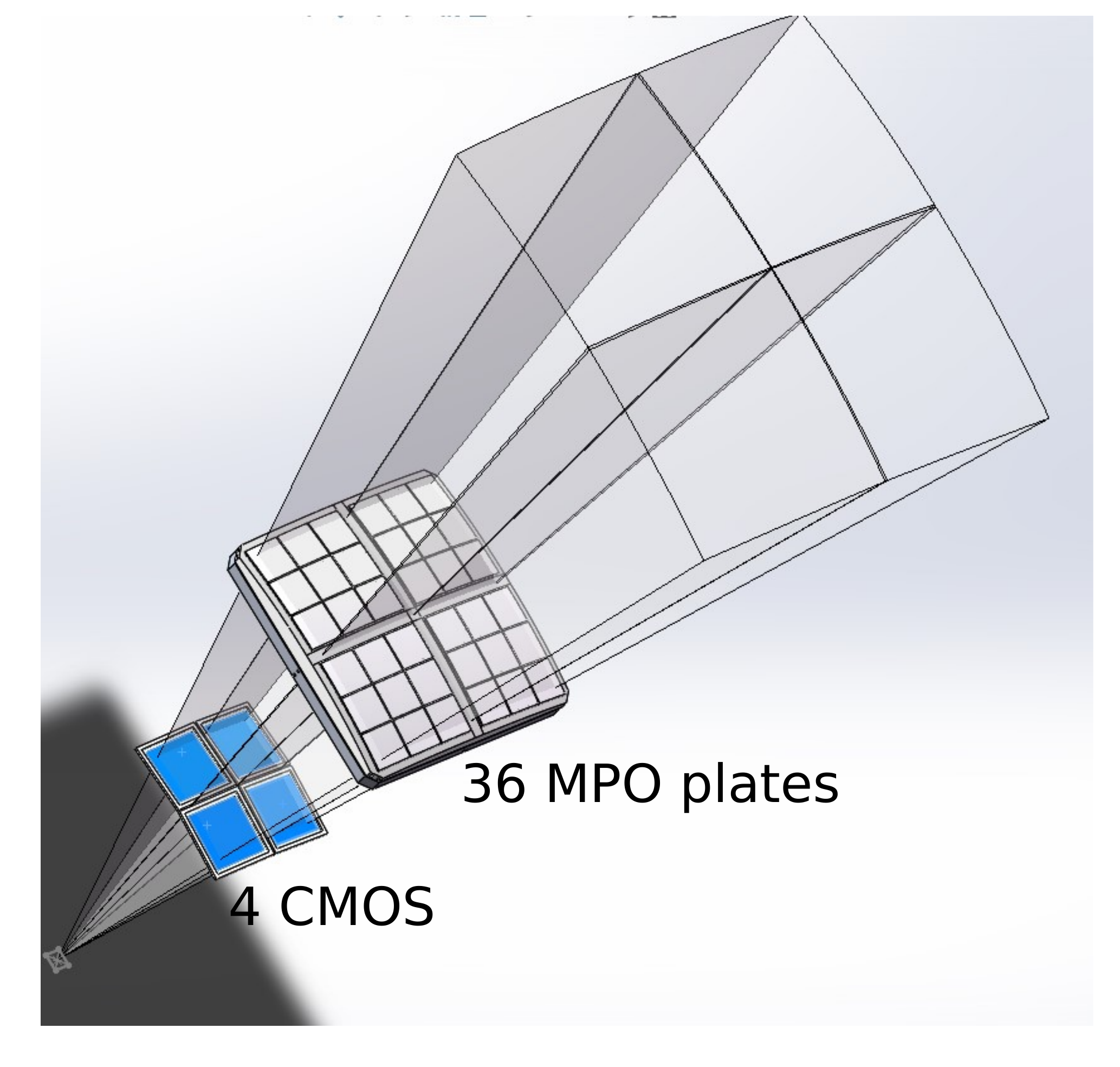}{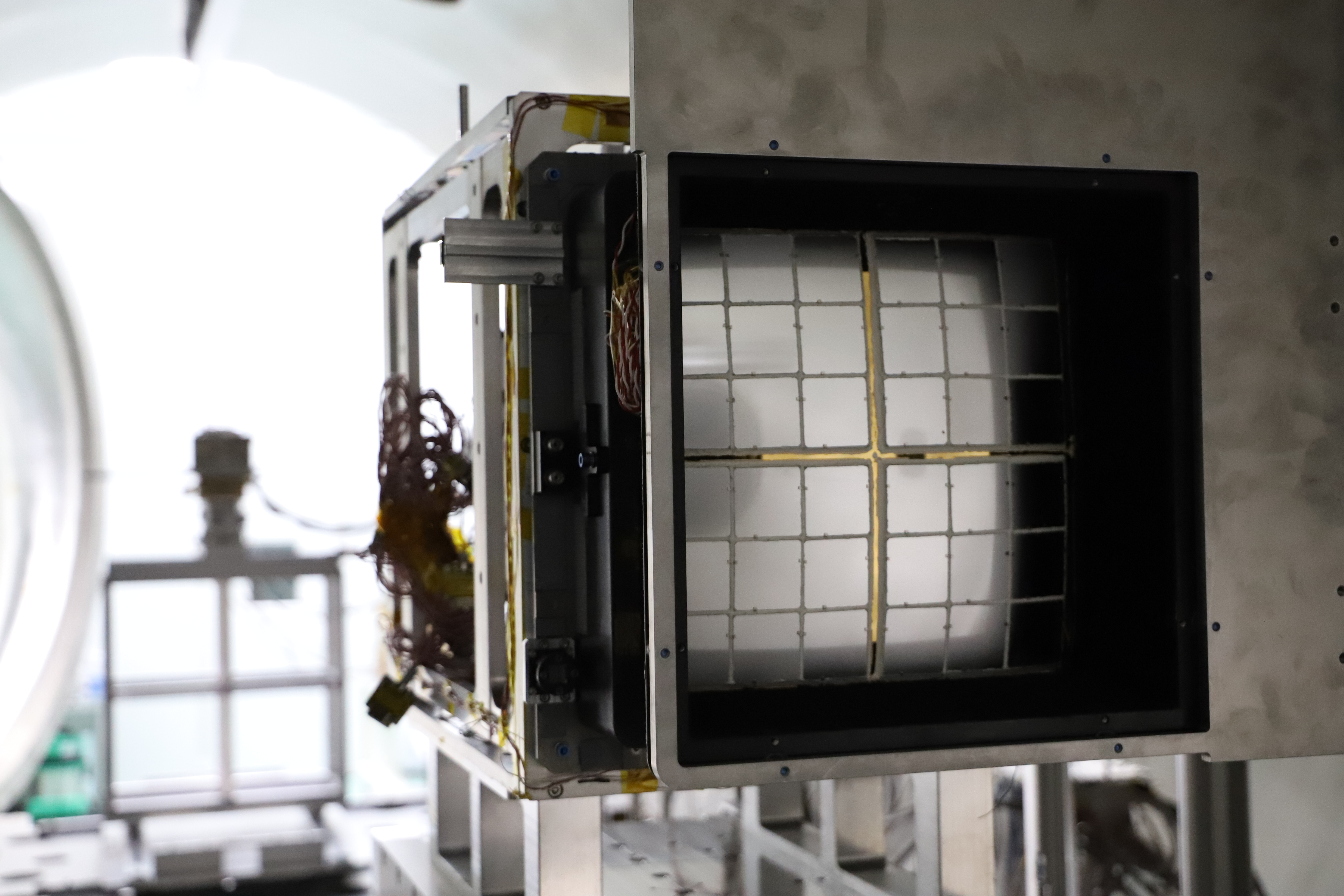}
\caption{Left: illustration of the configuration of the focusing mirror system, focal detector array and FoV of \lext. The mirror assembly is divided into 4 individual quadrants, each consisting of $3\times 3$ MPO plates and associated with one of the 4 detectors. The overall FoV of the telescope module is $18.6^\circ \times 18.6^\circ$. Right: a picture of the \lext instrument undergoing on-ground X-ray calibration at IHEP before being assembled onto the SATech satellite.}
\label{fig:lex}
\end{figure*}

\lext is a fully representative test model of one of the 12 identical modules of EP-WXT.
The design of EP-WXT is described in \citet{2018SPIE10699E..25Y,2022arXiv220909763Y} and Ling et al. in detail (in prep.) and is briefly summarized here. 
The EP-WXT module is composed of an imaging system, electronics, thermal control and mechanical structure. 
Its main specifications are given in Table \ref{tab:lext}.
The imaging system includes a mirror assembly (MA) and a focal detector array (Figure \ref{fig:lex}, left).
Based on the LE optics, the mirror is built from a mosaic of $6\times 6$ MPO plates slumped into spherically curved shape with a curvature radius of 750 mm.
The reflective surfaces of the pores are coated with Iridium. 
The MPO plates are individually mounted onto the supporting frame made of alloy. 
The mirror assembly is divided into four separate quadrants to take into account the gaps between the four detectors. The adjacent quadrants have an overlapping FoV of about 10 arcmin.
Each quadrant consists of $3\times 3$  MPO plates
and subtends a solid angle of $9.3^\circ \times 9.3^\circ$, defining its FoV.
The four make up the continuous overall FoV of $\sim 18.6^\circ \times 18.6^\circ$ of one of the EP-WXT modules.

The detector array is composed of four back-illuminated, large-format 
%(4\,k$\times$4\,k pixels,  6\,cm $\times$ 6\,cm in size) 
CMOS sensors, which
are mounted on the co-centering focal sphere with a radius half that of the MPOs.
The principle of the application of the CMOS to X-ray detection and imaging is similar to that of the traditional CCD sensors. 
Compared to CCD,  CMOS sensors have some advantages, such as fast readout speed (frame rate), relatively high operating temperature and thus relaxed cooling requirements, better radiation hardness and lower cost. 
Specifically, the CMOS sensors of \lext has a readout noise around 4\,e$^{-}$,  dark current $\sim$0.1 e$^{-}$\,s$^{-1}$\,pix$^{-1}$ at a temperature of -30\,$^{\circ}$C, a frame rate of 20 Hz, and 
an energy resolution of $\sim$130\,eV at 1.25 keV.
Some of the main parameters of the CMOS sensors are listed in Table \ref{tab:lext}, and the X-ray test results are summarized in \citet{2022PASP..134c5006W}.

The MPO and CMOS devices adopted were tested extensively via experiments at the National Astronomical Observatories, CAS (NAOC) and also at the University of Leicester independently \citep{2020charly_wxt_LU}.
The predicted performance of the designed LE telescope was studied by Monte-Carlo simulations, including the PSF, effective area, and predicted background spectrum \citep{2014SPIE.9144E..4EZ,2017ExA....43..267Z}.
% the sentences in previous are moved to the last paragraph of Sect.2.2
The mirror assembly was designed and built at the X-ray Imaging Lab (XIL) of NAOC, and the overall module (Figure \ref{fig:lex}, right) was designed and engineered at the Shanghai Institute of Technical Physics, CAS, and subsequently passed a series of space qualification tests.

\begin{deluxetable*}{lc}
%\tabletypesize{\scriptsize}
\tablewidth{0pt} 
%\tablenum{1}
\tablecaption{Specifications and performance of the instrument and SATech \label{tab:lext}}
\tablehead{
\colhead{Device/parameter} & \colhead{Value}
} 
%\colnumbers
\startdata 
Number of MPO plates & 36 \\
Size of MPO plate (mm) & $42.5 \times 42.5 \times 2.5 $ \\
Size of pore ($\micron$) & $40 \times40 $ \\
Focal length (mm) & 376.8$\pm 1.1$ \\
Field of view & $18.6^\circ \times 18.6^\circ$  \\
Angular resolution (arcmin) & 3.8--7.5, $\le 5$ (85\%) \\ 
Effective area$^a$ (cm$^2$) &  2--3 @1\,keV \\ 
Imaging sensors & 4 CMOS sensors\\
CMOS dimensions (mm) & $60 \times 60 $ \\
Pixel size ($\micron$), number & 15, $4{\rm k} \times 4{\rm k}$\\
Bandpass (keV) & 0.5--4.5 \\
Energy resolution (eV)  & 130  @1.25\,keV  \\
Readout speed (ms)  & 50 \\  
Payload mass (kg)  & 26 + 27 (electronics) \\
%Payload dimensions (cm) & $50? \times ? \times ? $ \\
Satellite mass (kg)  & 620 \\
Orbit & Sun-synchronous \\
Altitude (km) & 500 \\
Orbital period (min) & 95 \\
pointing accuracy (deg.) & 0.1 \\
Atitude stability (deg.\,s$^{-1}$) & 0.003 \\
Designed lifetime (years) & 2 \\ 
\enddata
\tablecomments{The values were measured from the on-ground calibrations. a) The effective area is for the central spot of the PSF.}
\end{deluxetable*}

\subsection{On-ground characterization}
%\subsection{On-ground calibration and characterization}
\label{sec:ong}

To fully characterize the actual performance of \lext, its key components and the complete module have been calibrated at several facilities.
The mirror assembly was calibrated at the PANTER facility of MPE \citep{2019VadiamXrayTest}  for the effective area, focus search and focal plane mapping, and independently at the X-ray Imaging Beamline (XIB) of NAOC \citep{2012zhangSPIE} for PSF and positioning accuracy. 
The four CMOS sensors were calibrated at NAOC for energy response with several characteristic lines of different elements.
The calibration of the complete module was performed at the 100m X-ray test Facility (100XF) \citep{Zhao2019Experiments,Wang2022IHEP100} at the Institute of High Energy Physics (IHEP), CAS, where the effective area, PSF and source positioning accuracy were calibrated at different incident angles and several energies of X-ray characteristic lines. 
The results of these calibrations will be presented elsewhere and only the basic characterizations are summarized here.

\begin{figure*}[htb]
\plottwo{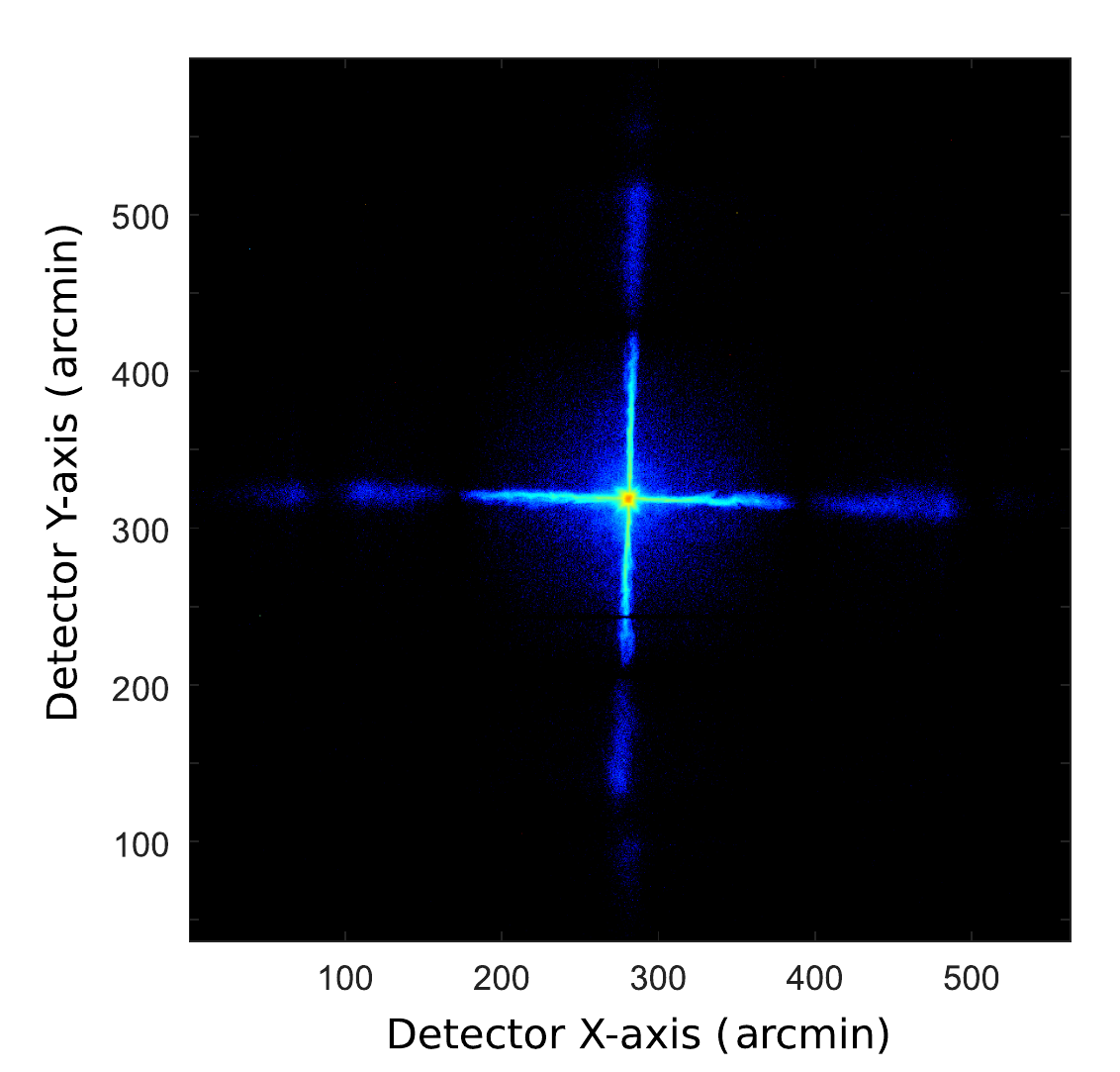}{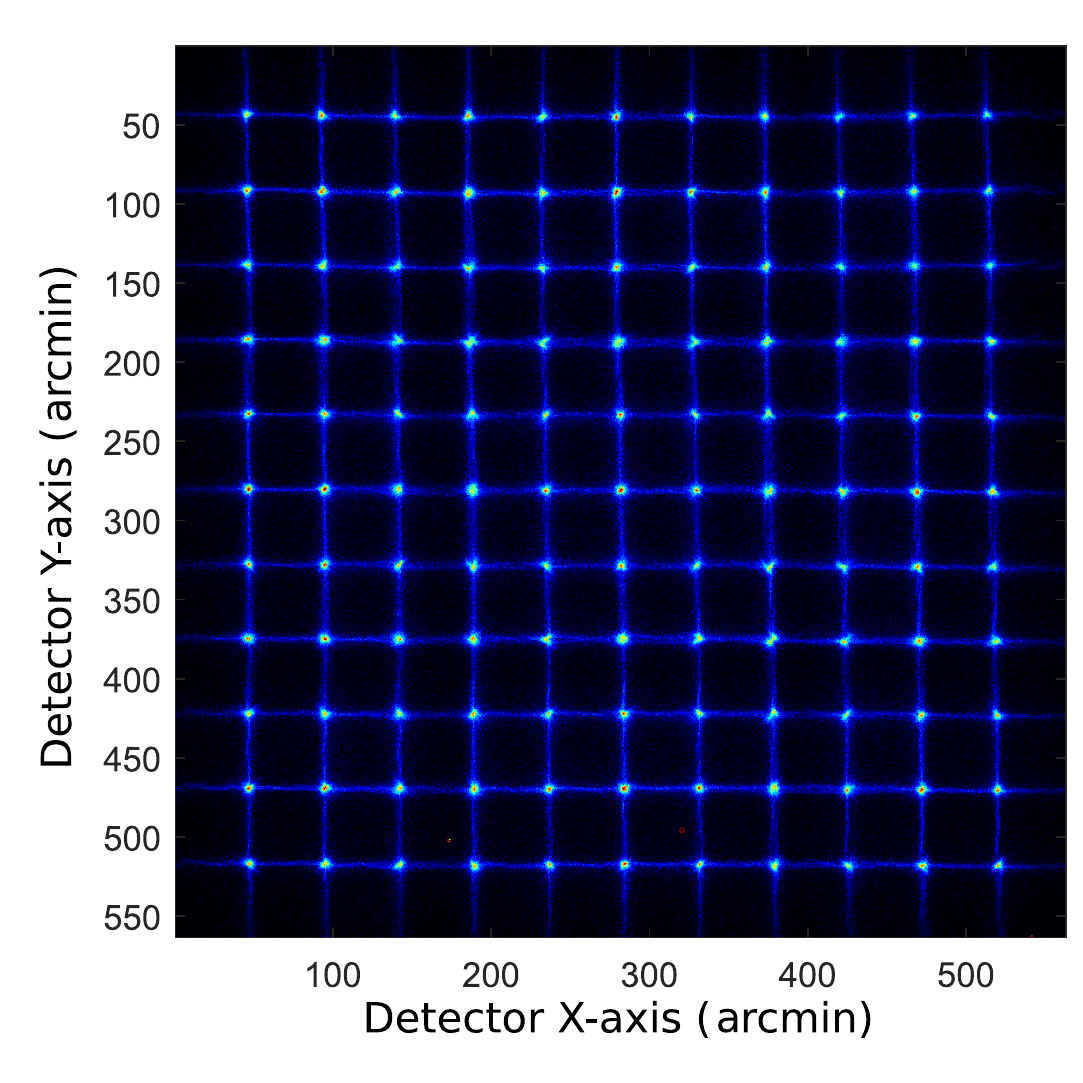}
\plottwo{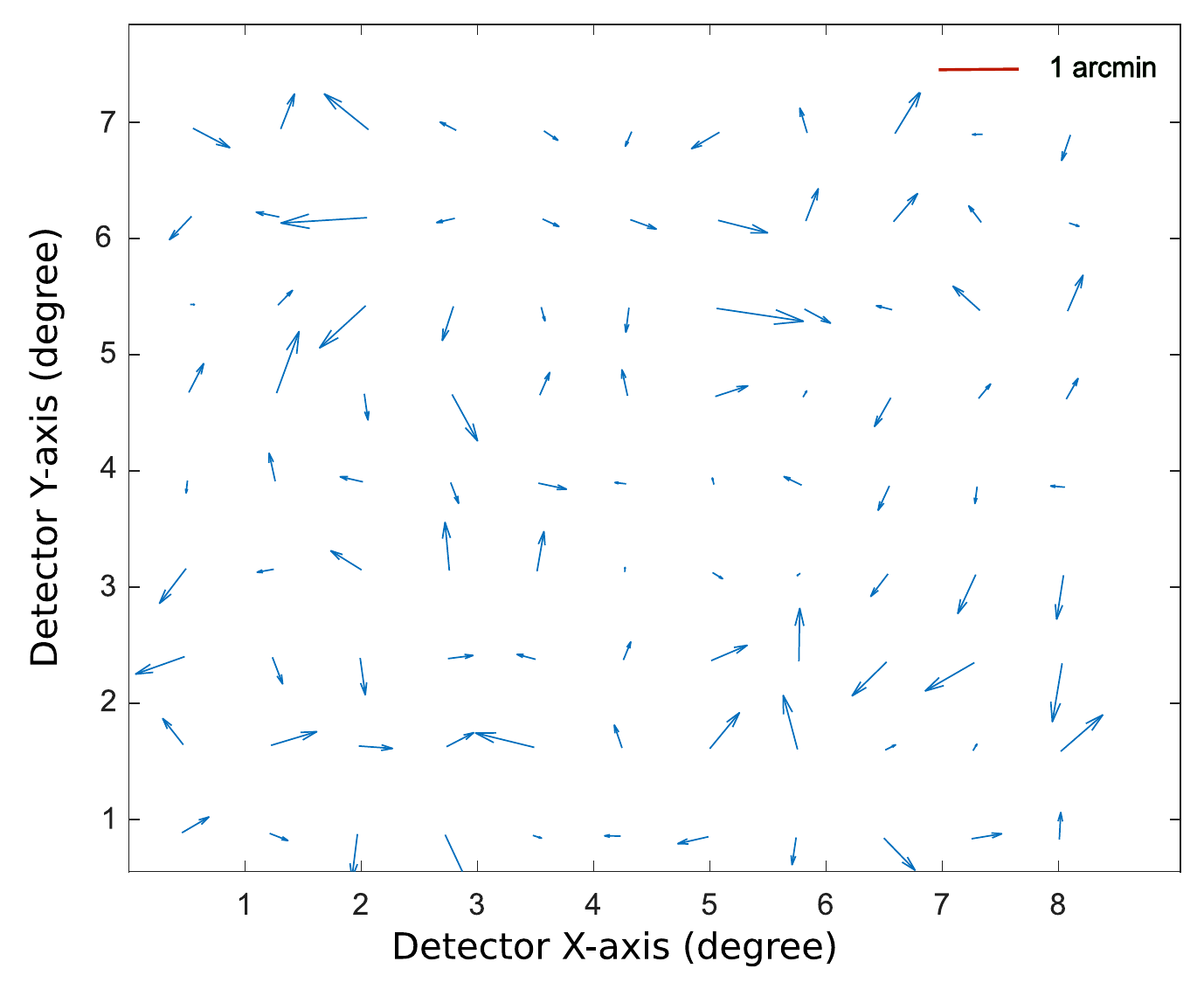}{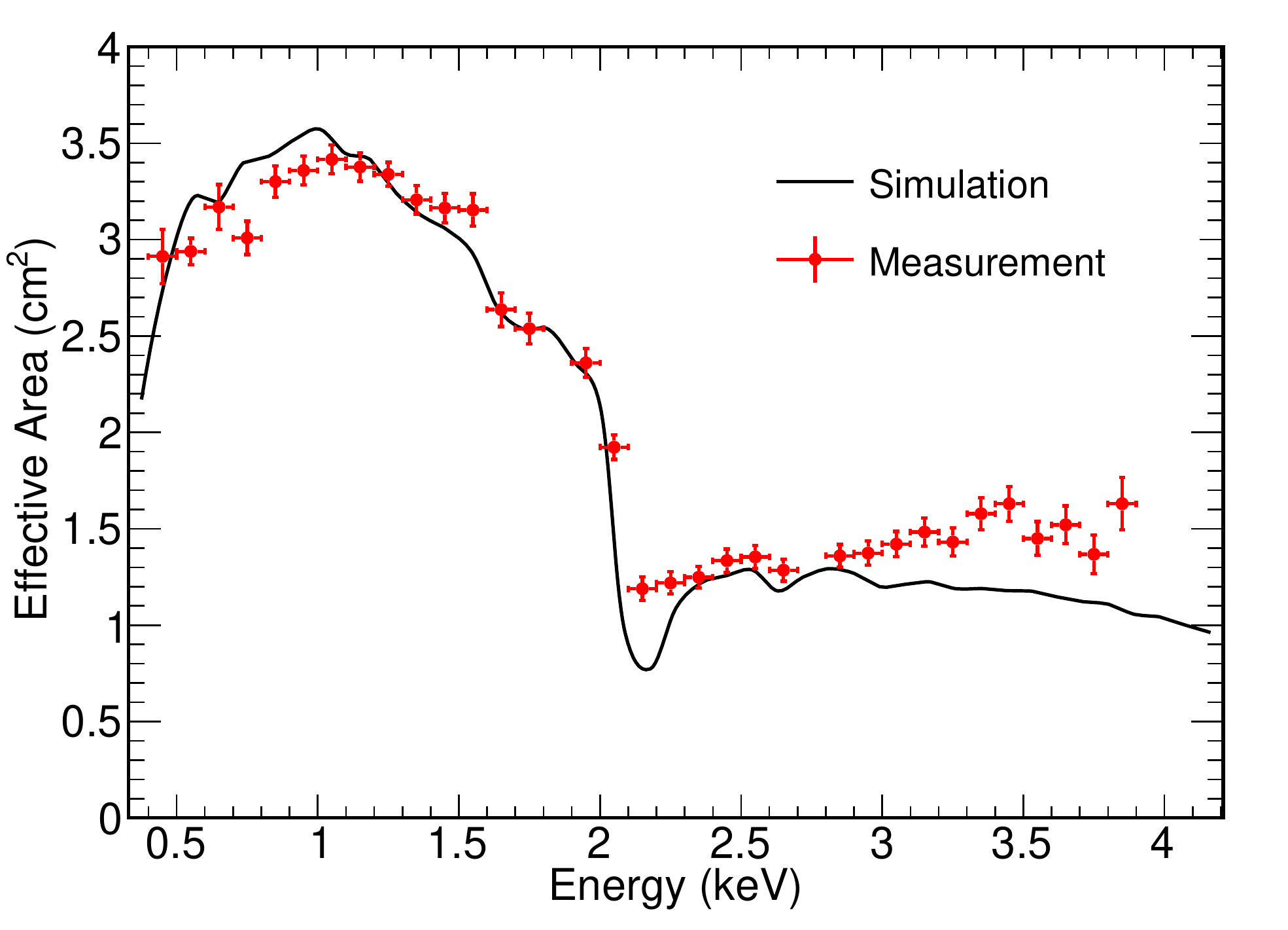}
\caption{Upper left: image of a point-like source (PSF) with an X-ray continuum spectrum peaking at 2.5\,keV seen at the center of a quadrant of the FoV obtained at XIB/NAOC (on a logarithmic color scale). 
Note that the image distance was adjusted accordingly for the finite X-ray source distance.   
Upper right: a mosaic of images of the PSF obtained at 1.25\,keV in $11\times 11$ directions across a quadrant of the FoV, taken with the complete module at IHEP/100XF. 
Lower left: positional deviations of the PSF centers from the source directions on the detector plane measured for a quadrant of the mirror assembly at NAOC/XIB. The maximum deviation is 1.1 arcmin. 
Lower right: a typical effective area curve of the mirror assembly for the central focal spot versus photon energy measured at MPE/PANTER and comparison with the simulation.}
\label{fig:cal}
\end{figure*}

A series of imaging scans was performed to measure the PSF and angular resolution of the MA in a grid of directions sampled uniformly across the entire FoV.  
A typical example of the measured PSF is shown in Figure \ref{fig:cal} (upper left), which composes a bright central spot and two cruciform arms, characteristic of the lobster eye optics.
These measurements well match the simulations \citep{2014SPIE.9144E..4EZ,2017ExA....43..267Z}.
The upper right panel of Figure \ref{fig:cal} shows mosaics of the X-ray images of a point-like source in $11\times 11$ directions across the FoV of one MA quadrant as an example. 
Within the entire module FoV, the PSFs show very similar characteristic shapes among all the sampled directions. 
The measured FWHMs of the central spot\footnote{Here we use the equivalent radius of a circle with the same area enclosed by the ellipse fitted to the contours of the 2D FWHM.} are in the range of 3.8--7.5 arcmin, with 5 arcmin at the 85th percentile. 
Positional deviations of the central PSF from the nominal direction of the source on the detector plane were also mapped (lower left panel of Figure \ref{fig:cal}), to calibrate the transform matrix of the detector coordinate system to the corresponding incidence angles of sources. 
The maximum deviation is 1.1 arcmin.
The measured effective areas at 1\,keV for the focal spot are in the range of 2--3 cm$^2$ across the FoV except at the edges. 
As an example, Figure \ref{fig:cal} (lower right) shows the effective area of the mirror assembly for one typical direction as a function of photon energy measured in the on-ground calibration.

When the two cruciform arms of the PSF are also taken into account, the effective area is $\sim 3$ times that for the central focal spot only. 
The result of the effective area calibration is the subject of a separate paper and will be presented elsewhere (Zhao et al. in prep.).
The measured PSF and effective areas agree largely with the simulation that was developed based mainly on Geant4 \citep{2017ExA....43..267Z}. 
Based on the above calibration results, a calibration database (CALDB) was built, which will be applied to the reduction and analysis of \lext data.

%These results demonstrate experimentally the prediction of the uniform, largely un-vignetted FoV of a LE imaging system.

\section{Flight experiment and results}
\label{sec:exp}

\subsection{Flight experiment}

\lext (Figure \ref{fig:lex}, right) is one of the experimental instruments aboard the SATech-01 satellite of the CAS,  which was launched on July 27, 2022.
Developed by the CAS's Innovation Academy for Microsatellites, SATech-01 is an exploration satellite aimed at test and demonstration of the new technologies of some 16 scientific experiments, 
ranging from astrophysics and solar physics, Earth observation, to space environment monitoring.
The satellite, with a designed lifetime of 2 years, is in a Sun-synchronous near-Earth circular orbit with a period of 95 minutes. 
Some of the satellite's key parameters are summarized in Table \ref{tab:lext}.
%The pointing accuracy of the spacecraft is 0.1 degree and the stability is 0.003 degree per second.
The precise pointing attitude of the satellite during observations is provided by two star trackers onboard.  
The satellite passes through the radiation belt at high geolatitude regions in each orbit and occasionally goes through the South Atlantic Anomaly (SAA). 
During these passages the detectors suffer from high backgrounds caused by charged particles.
The satellite can response to observations of Target of Opportunity (ToO) by uplinking commands via the S-band Telemetry/Command system with a latency of $\sim 24$ hours, or several hours for time-critical observations. 
The scientific data and house-keeping data are transmitted via the X-band telemetry to ground stations every day and then to the CAS's National Space Science Center, where the telemetry data are decoded, unpacked and verified. 
Afterwards, the data are sent to the EP Science Center (EPSC) at NAOC for reduction and analysis.
A detailed description of the SATech-01 satellite and \lext is to be presented elsewhere (Ling et al. in prep.).
 
In August and September 2022, \lext carried out a series of test observations for several days as part of its performance verification phase.
A number of pre-selected sky regions and targets were observed, including the Galactic Center, the Magellanic Clouds, Sco\,X-1, Cas A, Cyg Loop, and a few extragalactic sources.  
The observations were performed in Earth shadow to eliminate the effects of the Sun, starting 2 minutes after the satellite entering the shadow and ending 10 minutes before leaving it, resulting in an observational duration of $\sim$23 minutes in each orbit.  
The CMOS detectors were operating in the event mode.

\subsection{Data analysis}
\label{sec:data}
The telemetry data were converted into the standard  FITS format 
%\citep{2010A&A...524A..42P} 
and processed by a software pipeline developed for EP-WXT data reduction at EPSC.
The X-ray events were calibrated using the on-ground CALDB \footnote{In-orbit calibration has to await until the Crab nebula, the standard calibration source, is visible due to the Sun-avoidance constraint.} aforementioned.
A description of the data reduction and science products for EP-WXT is to be presented elsewhere (Liu et al. in prep.), and is only briefly summarized here. 
The data reduction algorithm for CMOS detectors is similar to that for CCDs, which are widely used for X-ray missions such as XMM-Newton and Swift/XRT.    

The bias residual is subtracted from each event and bad/flaring pixels are flagged. X-ray events are extracted with a grade and a pulse height amplitude assigned.    
The Pulse Invariant value of each event is calculated using the calibrated gain values. The position of each event is converted to celestial coordinates (J2000), 
using the relevant coordinate transform matrices in the CALDB.  
Single-, double-, triple-, and quadruple-events without anomalous flags are selected for further processing. 

To remove orbital intervals suffering from high particle backgrounds, 
a threshold on the geomagnetic cutoff rigidity (COR)$>5\,{\rm GV}$ is adopted. 
The orbital intervals passing through the SAA are also excluded.
The Earth elevation angle is set to $>10$ degrees.
These effects reduce the usable observing time to typically 6-15 minutes for each orbit. 

A cleaned events file is generated, with an exposure map that accounts for bad pixels and columns, attitude variations, and the distribution of the effective area across the FoV.
An image in the 0.5--4.0\,keV range is accumulated from the cleaned events, on which source detection is performed. The light curve and spectrum of each source found are extracted. The pipeline also generates the corresponding response matrix and ancillary response files to account for the distribution of the effective area and for PSF correction.
The data reduction is done for each of the detectors separately, and the resulting event files and images can further be merged.

\subsection{Results}
Here we report the most representative examples from the initial results, emphasizing  the wide field nature of the lobster eye optics. Detailed analysis of these and more observations will be presented in forthcoming papers.
It is found that the levels of the diffuse X-ray sky background, which dominates the energy band $< 2$ keV, agree generally with our simulation \citep{2017ExA....43..267Z}.
The particle background during the usable observational duration is $\sim10$ ${\rm cts\,s}^{-1}$ per CMOS in 0.5--4\,keV, which is twice the simulated value. 

\begin{figure*}[htb]
\plotone{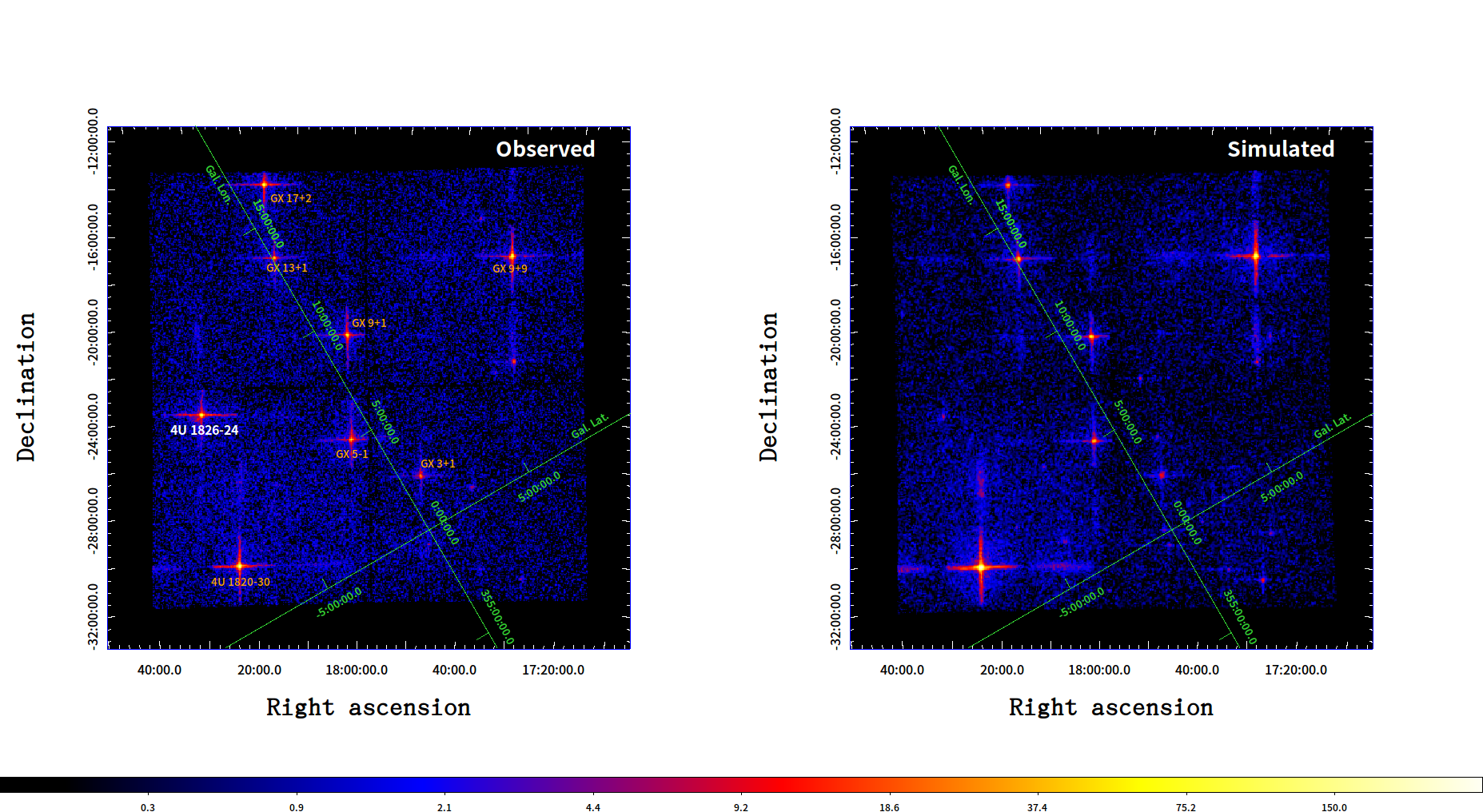}
\caption{First light X-ray image of the Galactic center region obtained by \lext in a one-shot observation of 798\,s in 0.5--4\,keV, covering a field of view $18.6^\circ \times 18.6^\circ$ (left). The simulated observation of the same patch of sky is also shown for comparison (right). The observation identifies a source (4U 1826-24) that had become obviously much brighter than viewed at a previous observation. Colors represent counts per pixel.
\label{fig:gc}}
\end{figure*}

Figure \ref{fig:gc} (left) shows the X-ray image of the Galactic Center region (centering on $l=6.6^\circ, b=0.9^\circ$), observed in one pointing with a net exposure of 798\,s on August 10, 2022. 
Within the FoV of $18.6^\circ \times 18.6^\circ$, 14 sources are detected at significance levels $\geq 5\,\sigma$ in one snapshot.
This is the first wide-field X-ray image of celestial bodies ever taken by a focusing imaging telescope.
The brightest sources are identified with the known X-ray binaries, including GX~9+9, GX~3+1, and 4U~1820-30.
Some fainter sources at flux levels $\sim1.0\times 10^{-10}$ \ergs\ ($\sim 3$\,mCrab) are also detected, e.g. 4U~1724-30 and 4U~1730-220. 
This flux is already below the one-day sensitivity of MAXI $\sim 15$\,mCrab \citep{2011PASJ...63S.635S}. 
Hence \lext is able to monitor the variability or outbursts of relatively faint sources on timescales as short as a thousand seconds, that are elusive for the previous and other ASMs in orbit.

We performed simulations to predict observed results based on previous surveys as input, including the ROSAT all-sky survey (RASS) \citep{RASS2016A&A...588A.103B} and MAXI \citep{MAXI2009PASJ...61..999M},
using an EP-WXT simulator which incorporates the on-ground CALDB. 
The observed image can be compared with the simulated one (Figure \ref{fig:gc}, right) with the same exposure. 
The similarity between the two images is striking. 
Interestingly, one source (4U~1826-24) is found to have brightened substantially compared to its previous flux used for the simulation, and an inspection into its long-term MAXI light curve\footnote{http://maxi.riken.jp/star\_data/J1829-237/J1829-237.html} confirmed this result. 
A preliminary analysis shows that the observed PSFs, showing no significant variations across the FoV, are generally consistent with those measured in the on-ground calibrations, although a more quantitative comparison is currently hampered by the small amount of source counts collected. 

\begin{figure*}[htb]
\plottwo{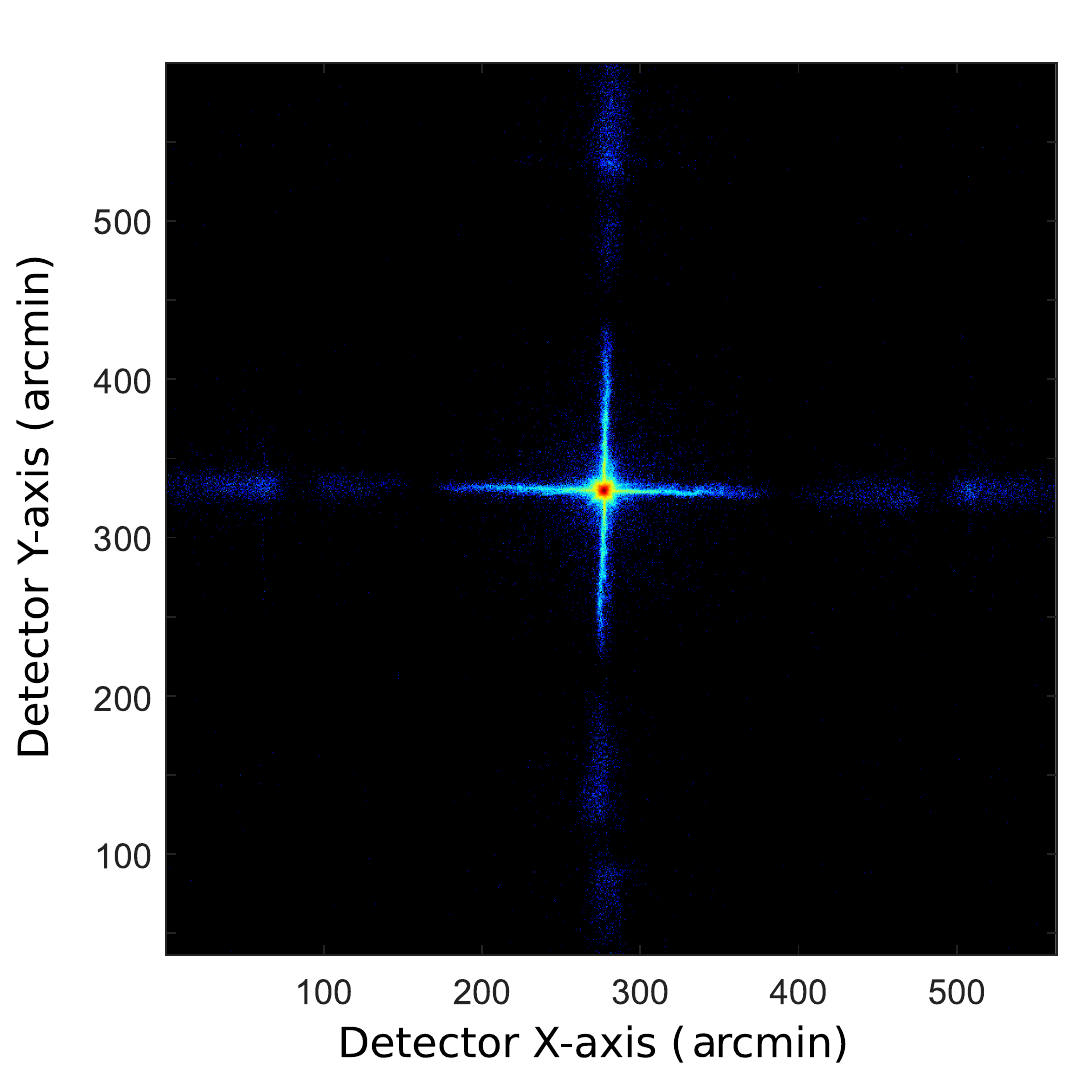}{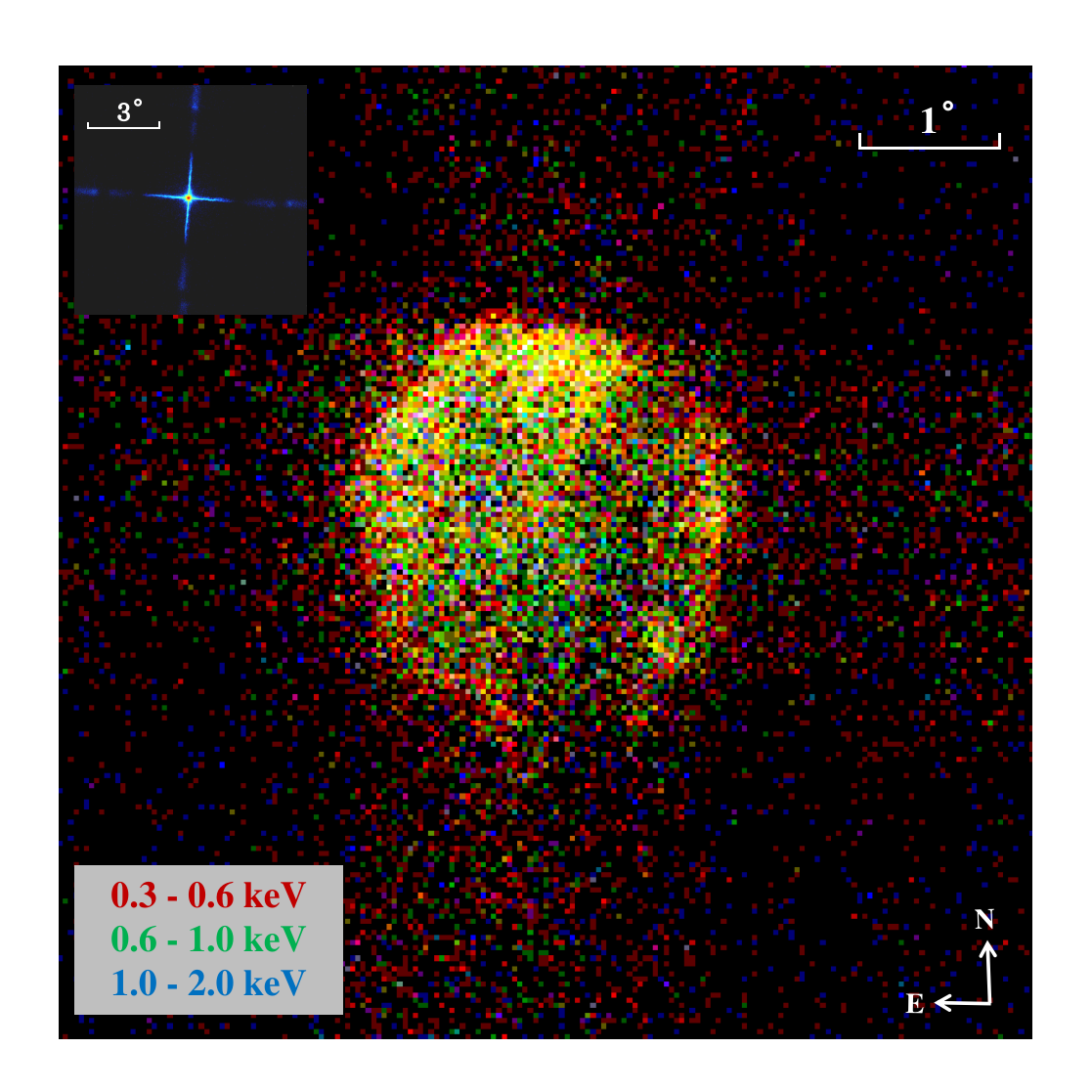}
\caption{
Left: X-ray image of Sco\,X-1 in 0.5--4\,keV observed by \lext  with 673\,s exposure, in excellent agreement with the PSF measured in on-ground calibration.  
Right: X-ray image of the Cygnus Loop nebula with a diameter of $\sim 2.5^\circ$ obtained with a 604\,s observation (colors represent photon energies). The inset shows the PSF measured from the observation of Sco\,X-1 (left).}
\label{fig:cl}
\end{figure*}

As a first demonstration of the in-flight imaging quality, \lext observed the brightest X-ray source, Sco\,X-1, on August 26, 2022.
The pile-up is negligible, thanks to the fast readout speed and small pixel size of the CMOS detectors (for a source as bright as 25\,Crab or 1000 counts\,s$^{-1}$, the pile-up fraction is $<1\%$).
The obtained image is shown in Figure \ref{fig:cl} (left), which was observed with 673\,s at the center of the same MA quadrant shown in Figure \ref{fig:cal} (upper left).
The observed count rate of Sco\,X-1 is $\sim 398.8$ cts\,s$^{-1}$ in 0.5--4 keV, corresponding to a flux of $\sim$10\,Crab.
With sufficient counts the PSF can be well sampled, and the measured FWHMs are 4.1 and 3.1 arcmin along the long and short axes of the PSF ellipse, consistent with the on-ground calibration result. 
We thus find no noticeable degradation of the imaging quality after launch. 
This also demonstrates \lext's capability of monitoring sources over a wide dynamic range of $10^4$ in flux. 

To demonstrate the imaging ability for extended sources on large-scale by the LE optics, 
Figure \ref{fig:cl} (right) shows the X-ray image of Cygnus Loop obtained with 604\,s exposure, in excellent agreement with the simulated image.
The overall structure is also well consistent with the images taken by other Wolter-I telescopes,
%, such as ROSAT \citep{ROSAT1999A&A...341..602A}, 
though the effect of the PSF's cruciform arms can still be seen.
%This demonstrates the imaging capability of a LE telescope for extended sources over a large scale. 

The quasar 3C\,382, a relatively faint X-ray source, was also observed for 606\,s and detected with 16 source counts within the PSF focal spot. 
The estimated flux limit is $(3-4)\times 10^{-11}$ \ergs\ for 1000\,s exposure at high Galactic latitude regions. 
This is 1.2--1.8 times the simulated value, but is broadly consistent with the expectations when the observed background level is adopted.

\section{Implications and conclusion}
\label{sec:con}

The initial results from the first in-flight experiments of \lext demonstrate that the observed images match well the simulations and the results of on-ground calibration. 
This has remarkable implications for both the technology and science of soft X-ray sky monitoring. 
A wide-field X-ray monitor based on the LE optics, as first proposed by \citet{1979ApJ...233..364A}, is now a concept proven to work in orbit.
The excellent agreements with the on-ground calibration indicate no noticeable degradation of the instrument performance during the launch and the first period of operation.
This relieves the long-standing concerns over the robustness of the MPO plates against launch, plates which are made of $\sim 2$\,mm-thin glass and largely hollow, and hence considered fragile. 

The measured PSF (around $\sim 5$\,arcmin in FWHM) and effective area (2--3 cm$^2$ ) show only mild variations across almost the entire FoV.
However, there are still deviations from the prediction of the uniform, un-vignetted FoV of an ideal, perfectly spherically symmetric LE optic.
These non-uniformities arise from a number of factors: the imperfectness introduced in the manufacturing and mounting of the optics, the mismatch between the flat detector plane and the spherical focal plane, and the obscuration of X-rays by the mounting frame between the individual MPO plates.
The first two factors affect mainly the point spread function, while the last one results in a smaller effective area in some directions than the nominal $\sim 3$ cm$^2$ of the MPO plates, causing  the non-uniformity as measured in the on-ground calibration (Zhao et al. in prep.).
When the above effects are considered, the measured instrumental properties are consistent with the prediction of a LE imaging system.

A combination of a wide, almost un-vignetted FoV and focusing imaging with several-arcmin resolution provides unprecedented capability and sensitivity of fast sky survey and monitoring. 
Even for a single module like \lext, the Grasp (FoV multiplied by effective area) reaches $\sim700$ sq.\,deg.\,cm$^2$ at 1 keV, almost twice the value for ROSAT and {\it XMM}-Newton. 
A LE ASM is capable of performing fast surveys over a large sky area by either mosaicking pointed snapshots or high-cadence scanning. 
The improved sensitivity, several times $10^{-11}$ \ergs for 1000\,s exposure, will enable the detection of fast X-ray transients beyond the reach of the current non-focusing instruments, such as X-ray flashes, GRB orphan afterglows, fireball flash of novae \citep{2022Natur.605..248K}, even possibly GRBs at high redshifts. 
Transients having fainter fluxes but longer timescales can be detected by stacking data of multiple observations, as the limiting flux scales inversely with square root of exposure time.    
For instance, as shown by simulations, a flux limit of  $\sim5\times10^{-12}$ \ergs can be reached with 50\,ks accumulated observations by \lext.
At this flux level, some of the known TDEs could be detected during their flaring phase, typically lasting from a few weeks to months. 
A considerable number of active galactic nuclei at a range of flux levels are also expected to be monitored at various timescales.
Of particular interest, \lext will serve as a novel instrument to search for potential counterparts of gravitational wave events in the soft X-ray band that is previously largely unexplored, during the upcoming operation runs of LIGO/VIRGO. 
This may be achieved by covering a considerable fraction of the probable locus of gravitational wave sources by the large FoV of \lext in one or a few snapshots.
This can be achieved by performing ToO observations by sending uplink commands with a latency from several up to 10 hours.

To conclude, snapshot images of the X-ray sky by a truly wide-field ($18.6^\circ \times 18.6^\circ $), grazing-incidence focusing telescope have been obtained for the first time. 
The first-light results from the \lext experiment mark the advent of the long-awaited wide-field LE X-ray telescopes. 
Preliminary analysis of the in-flight data shows excellent agreement on the results between the observed images and the on-ground calibration, as well as the simulations.
The experiment will lay a solid basis for the development of the present and proposed wide-field X-ray missions using LE MPO.
A combination of large FoV, order-of-magnitude increase in both sensitivity and angular resolution for X-ray ASM will ensure the promise of the science potential of X-ray sky monitoring, especially in the era of multi-wavelength and multi-messenger time-domain astronomy.

%% 
%% Also note that the acknowledgment environment does not support long amounts of text. If you have a lot of people and institutions to acknowledge, do not use this command. Instead, create a new \section{Acknowledgments}.
\begin{acknowledgments}
We thank all the members of the EP team, the EP consortium and the SATech-01 team.  This work is supported by the Einstein Probe project, a mission in the Strategic Priority Program on Space Science of CAS (Grants No.XDA15310000, XDA15052100). 
The CAS team acknowledge contribution from ESA for calibration of the mirror assembly and tests of part of the devices. 
This work was partially supported by International Partnership Program of CAS (Grant No.113111KYSB20190020). The Leicester and MPE teams acknowledge funding by ESA. The work performed at MPE's PANTER X-ray test facility has in part been supported by the European Union's Horizon 2020 Program under the AHEAD2020 project (Grant No. 871158). 
\end{acknowledgments}

\bibliography{lext.bib}

\begin{thebibliography}{}
\expandafter\ifx\csname natexlab\endcsname\relax\def\natexlab#1{#1}\fi
\providecommand{\url}[1]{\href{#1}{#1}}
\providecommand{\dodoi}[1]{doi:~\href{http://doi.org/#1}{\nolinkurl{#1}}}
\providecommand{\doeprint}[1]{\href{http://ascl.net/#1}{\nolinkurl{http://ascl.net/#1}}}
\providecommand{\doarXiv}[1]{\href{https://arxiv.org/abs/#1}{\nolinkurl{https://arxiv.org/abs/#1}}}

\bibitem[{{Abbott} {et~al.}(2017){Abbott}, {Abbott}, {Abbott}, {Acernese},
  {Ackley}, {Adams}, {Adams}, {Addesso}, {Adhikari}, {Adya}, {Affeldt},
  {Afrough}, {Agarwal}, {Agathos}, {Agatsuma}, {Aggarwal}, {Aguiar}, {Aiello},
  {Ain}, {Ajith}, {Allen}, {Allen}, {Allocca}, {Aloy}, {Altin}, {Amato},
  {Ananyeva}, {Anderson}, {Anderson}, {Angelova}, {Antier}, {Appert}, {Arai},
  {Araya}, {Areeda}, {Arnaud}, {Arun}, {Ascenzi}, {Ashton}, {Ast}, {Aston},
  {Astone}, {Atallah}, {Aufmuth}, {Aulbert}, {AultONeal}, {Austin},
  {Avila-Alvarez}, {Babak}, {Bacon}, {Bader}, {Bae}, {Baker}, {Baldaccini},
  {Ballardin}, {Ballmer}, {Banagiri}, {Barayoga}, {Barclay}, {Barish},
  {Barker}, {Barkett}, {Barone}, {Barr}, {Barsotti}, {Barsuglia}, {Barta},
  {Bartlett}, {Bartos}, {Bassiri}, {Basti}, {Batch}, {Bawaj}, {Bayley},
  {Bazzan}, {B{\'e}csy}, {Beer}, {Bejger}, {Belahcene}, {Bell}, {Berger},
  {Bergmann}, {Bero}, {Berry}, {Bersanetti}, {Bertolini}, {Betzwieser},
  {Bhagwat}, {Bhandare}, {Bilenko}, {Billingsley}, {Billman}, {Birch},
  {Birney}, {Birnholtz}, {Biscans}, {Biscoveanu}, {Bisht}, {Bitossi}, {Biwer},
  {Bizouard}, {Blackburn}, {Blackman}, {Blair}, {Blair}, {Blair}, {Bloemen},
  {Bock}, {Bode}, {Boer}, {Bogaert}, {Bohe}, {Bondu}, {Bonilla}, {Bonnand},
  {Boom}, {Bork}, {Boschi}, {Bose}, {Bossie}, {Bouffanais}, {Bozzi},
  {Bradaschia}, {Brady}, {Branchesi}, {Brau}, {Briant}, {Brillet}, {Brinkmann},
  {Brisson}, {Brockill}, {Broida}, {Brooks}, {Brown}, {Brown}, {Brunett},
  {Buchanan}, {Buikema}, {Bulik}, {Bulten}, {Buonanno}, {Buskulic}, {Buy},
  {Byer}, {Cabero}, {Cadonati}, {Cagnoli}, {Cahillane}, {Calder{\'o}n
  Bustillo}, {Callister}, {Calloni}, {Camp}, {Canepa}, {Canizares}, {Cannon},
  {Cao}, {Cao}, {Capano}, {Capocasa}, {Carbognani}, {Caride}, {Carney},
  {Casanueva Diaz}, {Casentini}, {Caudill}, {Cavagli{\`a}}, {Cavalier},
  {Cavalieri}, {Cella}, {Cepeda}, {Cerd{\'a}-Dur{\'a}n}, {Cerretani},
  {Cesarini}, {Chamberlin}, {Chan}, {Chao}, {Charlton}, {Chase},
  {Chassande-Mottin}, {Chatterjee}, {Chatziioannou}, {Cheeseboro}, {Chen},
  {Chen}, {Chen}, {Cheng}, {Chia}, {Chincarini}, {Chiummo}, {Chmiel}, {Cho},
  {Cho}, {Chow}, {Christensen}, {Chu}, {Chua}, {Chua}, {Chung}, {Chung},
  {Ciani}, {Ciolfi}, {Cirelli}, {Cirone}, {Clara}, {Clark}, {Clearwater},
  {Cleva}, {Cocchieri}, {Coccia}, {Cohadon}, {Cohen}, {Colla}, {Collette},
  {Cominsky}, {Constancio}, {Conti}, {Cooper}, {Corban}, {Corbitt},
  {Cordero-Carri{\'o}n}, {Corley}, {Cornish}, {Corsi}, {Cortese}, {Costa},
  {Coughlin}, {Coughlin}, {Coulon}, {Countryman}, {Couvares}, {Covas}, {Cowan},
  {Coward}, {Cowart}, {Coyne}, {Coyne}, {Creighton}, {Creighton}, {Cripe},
  {Crowder}, {Cullen}, {Cumming}, {Cunningham}, {Cuoco}, {Dal Canton},
  {D{\'a}lya}, {Danilishin}, {D'Antonio}, {Danzmann}, {Dasgupta}, {Da Silva
  Costa}, {Dattilo}, {Dave}, {Davier}, {Davis}, {Daw}, {Day}, {De}, {DeBra},
  {Degallaix}, {De Laurentis}, {Del{\'e}glise}, {Del Pozzo}, {Demos}, {Denker},
  {Dent}, {De Pietri}, {Dergachev}, {De Rosa}, {DeRosa}, {De Rossi}, {DeSalvo},
  {de Varona}, {Devenson}, {Dhurandhar}, {D{\'\i}az}, {Di Fiore}, {Di
  Giovanni}, {Di Girolamo}, {Di Lieto}, {Di Pace}, {Di Palma}, {Di Renzo},
  {Doctor}, {Dolique}, {Donovan}, {Dooley}, {Doravari}, {Dorrington},
  {Douglas}, {Dovale {\'A}lvarez}, {Downes}, {Drago}, {Dreissigacker},
  {Driggers}, {Du}, {Ducrot}, {Dupej}, {Dwyer}, {Edo}, {Edwards}, {Effler},
  {Eggenstein}, {Ehrens}, {Eichholz}, {Eikenberry}, {Eisenstein}, {Essick},
  {Estevez}, {Etienne}, {Etzel}, {Evans}, {Evans}, {Factourovich}, {Fafone},
  {Fair}, {Fairhurst}, {Fan}, {Farinon}, {Farr}, {Farr}, {Fauchon-Jones},
  {Favata}, {Fays}, {Fee}, {Fehrmann}, {Feicht}, {Fejer}, {Fernandez-Galiana},
  {Ferrante}, {Ferreira}, {Ferrini}, {Fidecaro}, {Finstad}, {Fiori},
  {Fiorucci}, {Fishbach}, {Fisher}, {Fitz-Axen}, {Flaminio}, {Fletcher},
  {Fong}, {Font}, {Forsyth}, {Forsyth}, {Fournier}, {Frasca}, {Frasconi},
  {Frei}, {Freise}, {Frey}, {Frey}, {Fries}, {Fritschel}, {Frolov}, {Fulda},
  {Fyffe}, {Gabbard}, {Gadre}, {Gaebel}, {Gair}, {Gammaitoni}, {Ganija},
  {Gaonkar}, {Garcia-Quiros}, {Garufi}, {Gateley}, {Gaudio}, {Gaur},
  {Gayathri}, {Gehrels}, {Gemme}, {Genin}, {Gennai}, {George}, {George},
  {Gergely}, {Germain}, {Ghonge}, {Ghosh}, {Ghosh}, {Ghosh}, {Giaime},
  {Giardina}, {Giazotto}, {Gill}, {Glover}, {Goetz}, {Goetz}, {Gomes},
  {Goncharov}, {Gonz{\'a}lez}, {Gonzalez Castro}, {Gopakumar}, {Gorodetsky},
  {Gossan}, {Gosselin}, {Gouaty}, {Grado}, {Graef}, {Granata}, {Grant}, {Gras},
  {Gray}, {Greco}, {Green}, {Gretarsson}, {Groot}, {Grote}, {Grunewald},
  {Gruning}, {Guidi}, {Guo}, {Gupta}, {Gupta}, {Gushwa}, {Gustafson},
  {Gustafson}, {Halim}, {Hall}, {Hall}, {Hamilton}, {Hammond}, {Haney},
  {Hanke}, {Hanks}, {Hanna}, {Hannam}, {Hannuksela}, {Hanson}, {Hardwick},
  {Harms}, {Harry}, {Harry}, {Hart}, {Haster}, {Haughian}, {Healy}, {Heidmann},
  {Heintze}, {Heitmann}, {Hello}, {Hemming}, {Hendry}, {Heng}, {Hennig},
  {Heptonstall}, {Heurs}, {Hild}, {Hinderer}, {Hoak}, {Hofman}, {Holt}, {Holz},
  {Hopkins}, {Horst}, {Hough}, {Houston}, {Howell}, {Hreibi}, {Hu}, {Huerta},
  {Huet}, {Hughey}, {Husa}, {Huttner}, {Huynh-Dinh}, {Indik}, {Inta}, {Intini},
  {Isa}, {Isac}, {Isi}, {Iyer}, {Izumi}, {Jacqmin}, {Jani}, {Jaranowski},
  {Jawahar}, {Jim{\'e}nez-Forteza}, {Johnson}, {Johnson-McDaniel}, {Jones},
  {Jones}, {Jonker}, {Ju}, {Junker}, {Kalaghatgi}, {Kalogera}, {Kamai},
  {Kandhasamy}, {Kang}, {Kanner}, {Kapadia}, {Karki}, {Karvinen}, {Kasprzack},
  {Kastaun}, {Katolik}, {Katsavounidis}, {Katzman}, {Kaufer}, {Kawabe},
  {K{\'e}f{\'e}lian}, {Keitel}, {Kemball}, {Kennedy}, {Kent}, {Key}, {Khalili},
  {Khan}, {Khan}, {Khan}, {Khazanov}, {Kijbunchoo}, {Kim}, {Kim}, {Kim}, {Kim},
  {Kim}, {Kim}, {Kimbrell}, {King}, {King}, {Kinley-Hanlon}, {Kirchhoff},
  {Kissel}, {Kleybolte}, {Klimenko}, {Knowles}, {Koch}, {Koehlenbeck}, {Koley},
  {Kondrashov}, {Kontos}, {Korobko}, {Korth}, {Kowalska}, {Kozak},
  {Kr{\"a}mer}, {Kringel}, {Krishnan}, {Kr{\'o}lak}, {Kuehn}, {Kumar}, {Kumar},
  {Kumar}, {Kuo}, {Kutynia}, {Kwang}, {Lackey}, {Lai}, {Landry}, {Lang},
  {Lange}, {Lantz}, {Lanza}, {Lartaux-Vollard}, {Lasky}, {Laxen}, {Lazzarini},
  {Lazzaro}, {Leaci}, {Leavey}, {Lee}, {Lee}, {Lee}, {Lee}, {Lee}, {Lehmann},
  {Lenon}, {Leonardi}, {Leroy}, {Letendre}, {Levin}, {Li}, {Linker},
  {Littenberg}, {Liu}, {Lo}, {Lockerbie}, {London}, {Lord}, {Lorenzini},
  {Loriette}, {Lormand}, {Losurdo}, {Lough}, {Lousto}, {Lovelace}, {L{\"u}ck},
  {Lumaca}, {Lundgren}, {Lynch}, {Ma}, {Macas}, {Macfoy}, {Machenschalk},
  {MacInnis}, {Macleod}, {Maga{\~n}a Hernandez}, {Maga{\~n}a-Sandoval},
  {Maga{\~n}a Zertuche}, {Magee}, {Majorana}, {Maksimovic}, {Man}, {Mandic},
  {Mangano}, {Mansell}, {Manske}, {Mantovani}, {Marchesoni}, {Marion},
  {M{\'a}rka}, {M{\'a}rka}, {Markakis}, {Markosyan}, {Markowitz}, {Maros},
  {Marquina}, {Martelli}, {Martellini}, {Martin}, {Martin}, {Martynov},
  {Mason}, {Massera}, {Masserot}, {Massinger}, {Masso-Reid}, {Mastrogiovanni},
  {Matas}, {Matichard}, {Matone}, {Mavalvala}, {Mazumder}, {McCarthy},
  {McClelland}, {McCormick}, {McCuller}, {McGuire}, {McIntyre}, {McIver},
  {McManus}, {McNeill}, {McRae}, {McWilliams}, {Meacher}, {Meadors}, {Mehmet},
  {Meidam}, {Mejuto-Villa}, {Melatos}, {Mendell}, {Mercer}, {Merilh},
  {Merzougui}, {Meshkov}, {Messenger}, {Messick}, {Metzdorff}, {Meyers},
  {Miao}, {Michel}, {Middleton}, {Mikhailov}, {Milano}, {Miller}, {Miller},
  {Miller}, {Millhouse}, {Milovich-Goff}, {Minazzoli}, {Minenkov}, {Ming},
  {Mishra}, {Mitra}, {Mitrofanov}, {Mitselmakher}, {Mittleman}, {Moffa},
  {Moggi}, {Mogushi}, {Mohan}, {Mohapatra}, {Montani}, {Moore}, {Moraru},
  {Moreno}, {Morriss}, {Mours}, {Mow-Lowry}, {Mueller}, {Muir}, {Mukherjee},
  {Mukherjee}, {Mukherjee}, {Mukund}, {Mullavey}, {Munch}, {Mu{\~n}iz},
  {Muratore}, {Murray}, {Napier}, {Nardecchia}, {Naticchioni}, {Nayak},
  {Neilson}, {Nelemans}, {Nelson}, {Nery}, {Neunzert}, {Nevin}, {Newport},
  {Newton}, {Ng}, {Nguyen}, {Nichols}, {Nielsen}, {Nissanke}, {Nitz}, {Noack},
  {Nocera}, {Nolting}, {North}, {Nuttall}, {Oberling}, {O'Dea}, {Ogin}, {Oh},
  {Oh}, {Ohme}, {Okada}, {Oliver}, {Oppermann}, {Oram}, {O'Reilly}, {Ormiston},
  {Ortega}, {O'Shaughnessy}, {Ossokine}, {Ottaway}, {Overmier}, {Owen}, {Pace},
  {Page}, {Page}, {Pai}, {Pai}, {Palamos}, {Palashov}, {Palomba}, {Pal-Singh},
  {Pan}, {Pan}, {Pang}, {Pang}, {Pankow}, {Pannarale}, {Pant}, {Paoletti},
  {Paoli}, {Papa}, {Parida}, {Parker}, {Pascucci}, {Pasqualetti},
  {Passaquieti}, {Passuello}, {Patil}, {Patricelli}, {Pearlstone}, {Pedraza},
  {Pedurand}, {Pekowsky}, {Pele}, {Penn}, {Perez}, {Perreca}, {Perri},
  {Pfeiffer}, {Phelps}, {Piccinni}, {Pichot}, {Piergiovanni}, {Pierro},
  {Pillant}, {Pinard}, {Pinto}, {Pirello}, {Pitkin}, {Poe}, {Poggiani},
  {Popolizio}, {Porter}, {Post}, {Powell}, {Prasad}, {Pratt}, {Pratten},
  {Predoi}, {Prestegard}, {Prijatelj}, {Principe}, {Privitera}, {Prodi},
  {Prokhorov}, {Puncken}, {Punturo}, {Puppo}, {P{\"u}rrer}, {Qi}, {Quetschke},
  {Quintero}, {Quitzow-James}, {Raab}, {Rabeling}, {Radkins}, {Raffai}, {Raja},
  {Rajan}, {Rajbhandari}, {Rakhmanov}, {Ramirez}, {Ramos-Buades}, {Rapagnani},
  {Raymond}, {Razzano}, {Read}, {Regimbau}, {Rei}, {Reid}, {Reitze}, {Ren},
  {Reyes}, {Ricci}, {Ricker}, {Rieger}, {Riles}, {Rizzo}, {Robertson}, {Robie},
  {Robinet}, {Rocchi}, {Rolland}, {Rollins}, {Roma}, {Romano}, {Romel},
  {Romie}, {Rosi{\'n}ska}, {Ross}, {Rowan}, {R{\"u}diger}, {Ruggi}, {Rutins},
  {Ryan}, {Sachdev}, {Sadecki}, {Sadeghian}, {Sakellariadou}, {Salconi},
  {Saleem}, {Salemi}, {Samajdar}, {Sammut}, {Sampson}, {Sanchez}, {Sanchez},
  {Sanchis-Gual}, {Sandberg}, {Sanders}, {Sassolas}, {Sathyaprakash},
  {Saulson}, {Sauter}, {Savage}, {Sawadsky}, {Schale}, {Scheel}, {Scheuer},
  {Schmidt}, {Schmidt}, {Schnabel}, {Schofield}, {Sch{\"o}nbeck}, {Schreiber},
  {Schuette}, {Schulte}, {Schutz}, {Schwalbe}, {Scott}, {Scott}, {Seidel},
  {Sellers}, {Sengupta}, {Sentenac}, {Sequino}, {Sergeev}, {Shaddock},
  {Shaffer}, {Shah}, {Shahriar}, {Shaner}, {Shao}, {Shapiro}, {Shawhan},
  {Sheperd}, {Shoemaker}, {Shoemaker}, {Siellez}, {Siemens}, {Sieniawska},
  {Sigg}, {Silva}, {Singer}, {Singh}, {Singhal}, {Sintes}, {Slagmolen},
  {Smith}, {Smith}, {Smith}, {Somala}, {Son}, {Sonnenberg}, {Sorazu},
  {Sorrentino}, {Souradeep}, {Spencer}, {Srivastava}, {Staats}, {Staley},
  {Steinke}, {Steinlechner}, {Steinlechner}, {Steinmeyer}, {Stevenson},
  {Stone}, {Stops}, {Strain}, {Stratta}, {Strigin}, {Strunk}, {Sturani},
  {Stuver}, {Summerscales}, {Sun}, {Sunil}, {Suresh}, {Sutton}, {Swinkels},
  {Szczepa{\'n}czyk}, {Tacca}, {Tait}, {Talbot}, {Talukder}, {Tanner},
  {T{\'a}pai}, {Taracchini}, {Tasson}, {Taylor}, {Taylor}, {Tewari}, {Theeg},
  {Thies}, {Thomas}, {Thomas}, {Thomas}, {Thorne}, {Thorne}, {Thrane},
  {Tiwari}, {Tiwari}, {Tokmakov}, {Toland}, {Tonelli}, {Tornasi},
  {Torres-Forn{\'e}}, {Torrie}, {T{\"o}yr{\"a}}, {Travasso}, {Traylor},
  {Trinastic}, {Tringali}, {Trozzo}, {Tsang}, {Tse}, {Tso}, {Tsukada}, {Tsuna},
  {Tuyenbayev}, {Ueno}, {Ugolini}, {Unnikrishnan}, {Urban}, {Usman},
  {Vahlbruch}, {Vajente}, {Valdes}, {van Bakel}, {van Beuzekom}, {van den
  Brand}, {Van Den Broeck}, {Vander-Hyde}, {van der Schaaf}, {van Heijningen},
  {van Veggel}, {Vardaro}, {Varma}, {Vass}, {Vas{\'u}th}, {Vecchio},
  {Vedovato}, {Veitch}, {Veitch}, {Venkateswara}, {Venugopalan}, {Verkindt},
  {Vetrano}, {Vicer{\'e}}, {Viets}, {Vinciguerra}, {Vine}, {Vinet}, {Vitale},
  {Vo}, {Vocca}, {Vorvick}, {Vyatchanin}, {Wade}, {Wade}, {Wade}, {Walet},
  {Walker}, {Wallace}, {Walsh}, {Wang}, {Wang}, {Wang}, {Wang}, {Wang}, {Ward},
  {Warner}, {Was}, {Watchi}, {Weaver}, {Wei}, {Weinert}, {Weinstein}, {Weiss},
  {Wen}, {Wessel}, {We{\ss}els}, {Westerweck}, {Westphal}, {Wette}, {Whelan},
  {Whitcomb}, {Whiting}, {Whittle}, {Wilken}, {Williams}, {Williams},
  {Williamson}, {Willis}, {Willke}, {Wimmer}, {Winkler}, {Wipf}, {Wittel},
  {Woan}, {Woehler}, {Wofford}, {Wong}, {Worden}, {Wright}, {Wu}, {Wysocki},
  {Xiao}, {Yamamoto}, {Yancey}, {Yang}, {Yap}, {Yazback}, {Yu}, {Yu}, {Yvert},
  {Zadro{\.z}ny}, {Zanolin}, {Zelenova}, {Zendri}, {Zevin}, {Zhang}, {Zhang},
  {Zhang}, {Zhang}, {Zhao}, {Zhou}, {Zhou}, {Zhu}, {Zhu}, {Zimmerman},
  {Zucker}, {Zweizig}, {(LIGO Scientific Collaboration}, {Virgo Collaboration},
  {Burns}, {Veres}, {Kocevski}, {Racusin}, {Goldstein}, {Connaughton},
  {Briggs}, {Blackburn}, {Hamburg}, {Hui}, {von Kienlin}, {McEnery}, {Preece},
  {Wilson-Hodge}, {Bissaldi}, {Cleveland}, {Gibby}, {Giles}, {Kippen},
  {McBreen}, {Meegan}, {Paciesas}, {Poolakkil}, {Roberts}, {Stanbro},
  {Gamma-ray Burst Monitor}, {Savchenko}, {Ferrigno}, {Kuulkers}, {Bazzano},
  {Bozzo}, {Brandt}, {Chenevez}, {Courvoisier}, {Diehl}, {Domingo}, {Hanlon},
  {Jourdain}, {Laurent}, {Lebrun}, {Lutovinov}, {Mereghetti}, {Natalucci},
  {Rodi}, {Roques}, {Sunyaev}, {Ubertini}, \&
  {(INTEGRAL}}]{2017ApJ...848L..13A}
{Abbott}, B.~P., {Abbott}, R., {Abbott}, T.~D., {et~al.} 2017, \apjl, 848, L13,
  \dodoi{10.3847/2041-8213/aa920c}

\bibitem[{{Angel}(1979)}]{1979ApJ...233..364A}
{Angel}, J.~R.~P. 1979, \apj, 233, 364, \dodoi{10.1086/157397}

\bibitem[{{Boller} {et~al.}(2016){Boller}, {Freyberg}, {Tr{\"u}mper}, {Haberl},
  {Voges}, \& {Nandra}}]{RASS2016A&A...588A.103B}
{Boller}, T., {Freyberg}, M.~J., {Tr{\"u}mper}, J., {et~al.} 2016, \aap, 588,
  A103, \dodoi{10.1051/0004-6361/201525648}

\bibitem[{Bradshaw {et~al.}(2019)Bradshaw, Burwitz, Hartner, Pelliciari,
  Langmeier, Liao, Friedrich, Valsecchi, Barri{\`e}re, Collon, \&
  Vacanti}]{2019VadiamXrayTest}
Bradshaw, M., Burwitz, V., Hartner, G., {et~al.} 2019, in Optics for EUV,
  X-Ray, and Gamma-Ray Astronomy IX, ed. S.~L. O'Dell \& G.~Pareschi, Vol.
  11119, International Society for Optics and Photonics (SPIE), 1111916,
  \dodoi{10.1117/12.2531709}

\bibitem[{{Brunton} {et~al.}(1999){Brunton}, {Martin}, {Fraser}, \&
  {Feller}}]{1999NIMPA.431..356B}
{Brunton}, A.~N., {Martin}, A.~P., {Fraser}, G.~W., \& {Feller}, W.~B. 1999,
  Nuclear Instruments and Methods in Physics Research A, 431, 356,
  \dodoi{10.1016/S0168-9002(99)00263-6}

\bibitem[{{Bunce} {et~al.}(2020){Bunce}, {Martindale}, {Lindsay}, {Muinonen},
  {Rothery}, {Pearson}, {McDonnell}, {Thomas}, {Thornhill}, {Tikkanen},
  {Feldman}, {Huovelin}, {Korpela}, {Esko}, {Lehtolainen}, {Treis}, {Majewski},
  {Hilchenbach}, {V{\"a}is{\"a}nen}, {Luttinen}, {Kohout}, {Penttil{\"a}},
  {Bridges}, {Joy}, {Alcacera-Gil}, {Alibert}, {Anand}, {Bannister},
  {Barcelo-Garcia}, {Bicknell}, {Blake}, {Bland}, {Butcher}, {Cheney},
  {Christensen}, {Crawford}, {Crawford}, {Dennerl}, {Dougherty}, {Drumm},
  {Fairbend}, {Genzer}, {Grande}, {Hall}, {Hodnett}, {Houghton}, {Imber},
  {Kallio}, {Lara}, {Balado Margeli}, {Mas-Hesse}, {Maurice}, {Milan},
  {Millington-Hotze}, {Nenonen}, {Nittler}, {Okada}, {Orm{\"o}},
  {Perez-Mercader}, {Poyner}, {Robert}, {Ross}, {Pajas-Sanz}, {Schyns},
  {Seguy}, {Str{\"u}der}, {Vaudon}, {Viceira-Mart{\'\i}n}, {Williams},
  {Willingale}, \& {Yeoman}}]{2020SSRv..216..126B}
{Bunce}, E.~J., {Martindale}, A., {Lindsay}, S., {et~al.} 2020, \ssr, 216, 126,
  \dodoi{10.1007/s11214-020-00750-2}

\bibitem[{{Chapman} {et~al.}(1991){Chapman}, {Nugent}, \&
  {Wilkins}}]{1991RScI...62.1542C}
{Chapman}, H.~N., {Nugent}, K.~A., \& {Wilkins}, S.~W. 1991, Review of
  Scientific Instruments, 62, 1542, \dodoi{10.1063/1.1142432}

\bibitem[{{Chen} {et~al.}(2020){Chen}, {Cui}, {Han}, {Wang}, {Yang}, {Wang},
  {Li}, {Ma}, {Xu}, {Lu}, {Chen}, {Tang}, {Yuan}, {Friedrich}, {Meidinger},
  {Keil}, {Burwitz}, {Eder}, {Hartmann}, {Nandra}, {Keereman}, {Santovincenzo},
  {Vernani}, {Bianucci}, {Valsecchi}, {Wang}, {Wang}, {Wang}, {Li}, {Sheng},
  {Qiang}, {Shi}, {Chao}, {Song}, {Zhang}, {Huo}, {Wang}, {Cong}, {Yang},
  {Hou}, {Zhao}, {Zhao}, {Chen}, {Li}, {Zhang}, {Luo}, {Xu}, {Li}, {Zhang},
  {Bi}, {Zhu}, {Yu}, {Chen}, {Lv}, {Lu}, \& {Zhang}}]{2020SPIE11444E..5BC}
{Chen}, Y., {Cui}, W., {Han}, D., {et~al.} 2020, in Society of Photo-Optical
  Instrumentation Engineers (SPIE) Conference Series, Vol. 11444, Society of
  Photo-Optical Instrumentation Engineers (SPIE) Conference Series, 114445B,
  \dodoi{10.1117/12.2562311}

\bibitem[{{Collier} {et~al.}(2015){Collier}, {Porter}, {Sibeck}, {Carter},
  {Chiao}, {Chornay}, {Cravens}, {Galeazzi}, {Keller}, {Koutroumpa},
  {Kujawski}, {Kuntz}, {Read}, {Robertson}, {Sembay}, {Snowden}, {Thomas},
  {Uprety}, \& {Walsh}}]{2015RScI...86g1301C}
{Collier}, M.~R., {Porter}, F.~S., {Sibeck}, D.~G., {et~al.} 2015, Review of
  Scientific Instruments, 86, 071301, \dodoi{10.1063/1.4927259}

\bibitem[{Feldman {et~al.}(2020)Feldman, O'Brien, Willingale, Zhang, Ling,
  Yuan, Jia, Jin, Li, Xu, Zhang, Lerman, Hutchinson, McHugh, \&
  Lodge}]{2020charly_wxt_LU}
Feldman, C., O'Brien, P., Willingale, R., {et~al.} 2020, in Space Telescopes
  and Instrumentation 2020: Ultraviolet to Gamma Ray, ed. J.-W.~A. den Herder,
  S.~Nikzad, \& K.~Nakazawa, Vol. 11444, International Society for Optics and
  Photonics (SPIE), 114447R, \dodoi{10.1117/12.2562194}

\bibitem[{{Feldman}(2022)}]{2022SPIE121811P..7RF}
{Feldman}, Charlotte, e.~a. 2022, in Society of Photo-Optical Instrumentation
  Engineers (SPIE) Conference Series, Vol. 12181, Society of Photo-Optical
  Instrumentation Engineers (SPIE) Conference Series, 121811P

\bibitem[{{Fraser}(2009)}]{2009xrda.book.....F}
{Fraser}, G.~W. 2009, {X-ray Detectors in Astronomy}

\bibitem[{{Fraser} {et~al.}(1993){Fraser}, {Brunton}, {Lees}, {Pearson}, \&
  {Feller}}]{1993NIMPA.324..404F}
{Fraser}, G.~W., {Brunton}, A.~N., {Lees}, J.~E., {Pearson}, J.~F., \&
  {Feller}, W.~B. 1993, Nuclear Instruments and Methods in Physics Research A,
  324, 404, \dodoi{10.1016/0168-9002(93)91003-6}

\bibitem[{{Fraser} {et~al.}(1992){Fraser}, {Lees}, {Pearson}, {Sims}, \&
  {Roxburgh}}]{1992SPIE.1546...41F}
{Fraser}, G.~W., {Lees}, J.~E., {Pearson}, J.~F., {Sims}, M.~R., \& {Roxburgh},
  K. 1992, in Society of Photo-Optical Instrumentation Engineers (SPIE)
  Conference Series, Vol. 1546, Multilayer and Grazing Incidence X-Ray/EUV
  Optics, ed. R.~B. {Hoover}, 41--52, \dodoi{10.1117/12.51224}

\bibitem[{{Fraser} {et~al.}(2002){Fraser}, {Brunton}, {Bannister}, {Pearson},
  {Ward}, {Stevenson}, {Watson}, {Warwick}, {Whitehead}, {O'Brian}, {White},
  {Jahoda}, {Black}, {Hunter}, {Deines-Jones}, {Priedhorsky}, {Brumby},
  {Borozdin}, {Vestrand}, {Fabian}, {Nugent}, {Peele}, {Irving}, {Price},
  {Eckersley}, {Renouf}, {Smith}, {Parmar}, {McHardy}, {Uttley}, \&
  {Lawrence}}]{2002SPIE.4497..115F}
{Fraser}, G.~W., {Brunton}, A.~N., {Bannister}, N.~P., {et~al.} 2002, in
  Society of Photo-Optical Instrumentation Engineers (SPIE) Conference Series,
  Vol. 4497, X-Ray and Gamma-Ray Instrumentation for Astronomy XII, ed. K.~A.
  {Flanagan} \& O.~H.~W. {Siegmund}, 115--126, \dodoi{10.1117/12.454217}

\bibitem[{{Gehrels} \& {Cannizzo}(2015)}]{2015JHEAp...7....2G}
{Gehrels}, N., \& {Cannizzo}, J.~K. 2015, Journal of High Energy Astrophysics,
  7, 2, \dodoi{10.1016/j.jheap.2015.03.003}

\bibitem[{{Gehrels} {et~al.}(2004){Gehrels}, {Chincarini}, {Giommi}, {Mason},
  {Nousek}, {Wells}, {White}, {Barthelmy}, {Burrows}, {Cominsky}, {Hurley},
  {Marshall}, {M{\'e}sz{\'a}ros}, {Roming}, {Angelini}, {Barbier}, {Belloni},
  {Campana}, {Caraveo}, {Chester}, {Citterio}, {Cline}, {Cropper}, {Cummings},
  {Dean}, {Feigelson}, {Fenimore}, {Frail}, {Fruchter}, {Garmire}, {Gendreau},
  {Ghisellini}, {Greiner}, {Hill}, {Hunsberger}, {Krimm}, {Kulkarni}, {Kumar},
  {Lebrun}, {Lloyd-Ronning}, {Markwardt}, {Mattson}, {Mushotzky}, {Norris},
  {Osborne}, {Paczynski}, {Palmer}, {Park}, {Parsons}, {Paul}, {Rees},
  {Reynolds}, {Rhoads}, {Sasseen}, {Schaefer}, {Short}, {Smale}, {Smith},
  {Stella}, {Tagliaferri}, {Takahashi}, {Tashiro}, {Townsley}, {Tueller},
  {Turner}, {Vietri}, {Voges}, {Ward}, {Willingale}, {Zerbi}, \&
  {Zhang}}]{2004ApJ...611.1005G}
{Gehrels}, N., {Chincarini}, G., {Giommi}, P., {et~al.} 2004, \apj, 611, 1005,
  \dodoi{10.1086/422091}

\bibitem[{{G{\"o}tz} {et~al.}(2016){G{\"o}tz}, {Meuris}, {Pinsard},
  {Doumayrou}, {Tourrette}, {Osborne}, {Willingale}, {Sykes}, {Pearson}, {Le
  Duigou}, \& {Mercier}}]{2016SPIE.9905E..4LG}
{G{\"o}tz}, D., {Meuris}, A., {Pinsard}, F., {et~al.} 2016, in Society of
  Photo-Optical Instrumentation Engineers (SPIE) Conference Series, Vol. 9905,
  Space Telescopes and Instrumentation 2016: Ultraviolet to Gamma Ray, ed.
  J.-W.~A. {den Herder}, T.~{Takahashi}, \& M.~{Bautz}, 99054L,
  \dodoi{10.1117/12.2232484}

\bibitem[{{Holt} \& {Priedhorsky}(1987)}]{1987SSRv...45..269H}
{Holt}, S.~S., \& {Priedhorsky}, W. 1987, \ssr, 45, 269,
  \dodoi{10.1007/BF00171996}

\bibitem[{{Hudec} \& {Feldman}(2022)}]{2022arXiv220807149H}
{Hudec}, R., \& {Feldman}, C. 2022, arXiv e-prints, arXiv:2208.07149.
\newblock \doarXiv{2208.07149}

\bibitem[{{Hudec} {et~al.}(2017){Hudec}, {Sveda}, {P{\'\i}na}, {Inneman},
  {Semencova}, \& {Skulinova}}]{2017SPIE10567E..19H}
{Hudec}, R., {Sveda}, L., {P{\'\i}na}, L., {et~al.} 2017, in Society of
  Photo-Optical Instrumentation Engineers (SPIE) Conference Series, Vol. 10567,
  Society of Photo-Optical Instrumentation Engineers (SPIE) Conference Series,
  1056719, \dodoi{10.1117/12.2308126}

\bibitem[{{Kaaret} {et~al.}(1992){Kaaret}, {Geissbuehler}, {Chen}, \&
  {Glavinas}}]{1992ApOpt..31.7339K}
{Kaaret}, P., {Geissbuehler}, P., {Chen}, A., \& {Glavinas}, E. 1992, \ao, 31,
  7339, \dodoi{10.1364/AO.31.007339}

\bibitem[{{K{\"o}nig} {et~al.}(2022){K{\"o}nig}, {Wilms}, {Arcodia}, {Dauser},
  {Dennerl}, {Doroshenko}, {Haberl}, {H{\"a}mmerich}, {Kirsch}, {Kreykenbohm},
  {Lorenz}, {Malyali}, {Merloni}, {Rau}, {Rauch}, {Sala}, {Schwope},
  {Suleimanov}, {Weber}, \& {Werner}}]{2022Natur.605..248K}
{K{\"o}nig}, O., {Wilms}, J., {Arcodia}, R., {et~al.} 2022, \nat, 605, 248,
  \dodoi{10.1038/s41586-022-04635-y}

\bibitem[{{Matsuoka} {et~al.}(2009){Matsuoka}, {Kawasaki}, {Ueno}, {Tomida},
  {Kohama}, {Suzuki}, {Adachi}, {Ishikawa}, {Mihara}, {Sugizaki}, {Isobe},
  {Nakagawa}, {Tsunemi}, {Miyata}, {Kawai}, {Kataoka}, {Morii}, {Yoshida},
  {Negoro}, {Nakajima}, {Ueda}, {Chujo}, {Yamaoka}, {Yamazaki}, {Nakahira},
  {You}, {Ishiwata}, {Miyoshi}, {Eguchi}, {Hiroi}, {Katayama}, \&
  {Ebisawa}}]{MAXI2009PASJ...61..999M}
{Matsuoka}, M., {Kawasaki}, K., {Ueno}, S., {et~al.} 2009, \pasj, 61, 999,
  \dodoi{10.1093/pasj/61.5.999}

\bibitem[{{O'Brien} {et~al.}(2020){O'Brien}, {Hutchinson}, {Lerman}, {Feldman},
  {McHugh}, {Lodge}, {Willingale}, {Beardmore}, {Speight}, \&
  {Drumm}}]{2020SPIE11444E..2LO}
{O'Brien}, P., {Hutchinson}, I., {Lerman}, H., {et~al.} 2020, in Society of
  Photo-Optical Instrumentation Engineers (SPIE) Conference Series, Vol. 11444,
  Society of Photo-Optical Instrumentation Engineers (SPIE) Conference Series,
  114442L, \dodoi{10.1117/12.2561301}

\bibitem[{{Peele} {et~al.}(1996){Peele}, {Nugent}, {Rode}, {Gabel},
  {Richardson}, {Strack}, \& {Siegmund}}]{1996ApOpt..35.4420P}
{Peele}, A.~G., {Nugent}, K.~A., {Rode}, A.~V., {et~al.} 1996, \ao, 35, 4420,
  \dodoi{10.1364/AO.35.004420}

\bibitem[{{Priedhorsky} {et~al.}(1996){Priedhorsky}, {Peele}, \&
  {Nugent}}]{1996MNRAS.279..733P}
{Priedhorsky}, W.~C., {Peele}, A.~G., \& {Nugent}, K.~A. 1996, \mnras, 279,
  733, \dodoi{10.1093/mnras/279.3.733}

\bibitem[{{Sembay} {et~al.}(2016){Sembay}, {Branduardi-Raymont}, {Drumm},
  {Escoubet}, {Genov}, {Gow}, {Hall}, {Holland}, {Hudec}, {Mas-Hesse},
  {Kennedy}, {Kuntz}, {Nakamura}, {Ostgaard}, {Ottensamer}, {Raab}, {Read},
  {Rebuffat}, {Romstedt}, {Schyns}, {Sibeck}, {Srp}, {Steller}, {Sun}, {Sykes},
  {Thornhill}, {Walsh}, {Walton}, {Wang}, {Wei}, {Wielders}, \&
  {Whittaker}}]{2016AGUFMSM44A..04S}
{Sembay}, S., {Branduardi-Raymont}, G., {Drumm}, P., {et~al.} 2016, in AGU Fall
  Meeting Abstracts, SM44A--04

\bibitem[{{Sugizaki} {et~al.}(2011){Sugizaki}, {Mihara}, {Serino}, {Yamamoto},
  {Matsuoka}, {Kohama}, {Tomida}, {Ueno}, {Kawai}, {Morii}, {Sugimori},
  {Nakahira}, {Yamaoka}, {Yoshida}, {Nakajima}, {Negoro}, {Eguchi}, {Isobe},
  {Ueda}, \& {Tsunemi}}]{2011PASJ...63S.635S}
{Sugizaki}, M., {Mihara}, T., {Serino}, M., {et~al.} 2011, \pasj, 63, S635,
  \dodoi{10.1093/pasj/63.sp3.S635}

\bibitem[{WANG {et~al.}(2022, in press)WANG, ZHAO, HOU, YANG, CHEN, LI, ZHU,
  ZHAO, MA, XU, CHEN, WANG, LU, ZHANG, ZHANG, CHEN, \& XU}]{Wang2022IHEP100}
WANG, Y.-s., ZHAO, Z.-j., HOU, D.-j., {et~al.} 2022, in press, Experimental
  Astronomy

\bibitem[{{Wilkins} {et~al.}(1989){Wilkins}, {Stevenson}, {Nugent}, {Chapman},
  \& {Steenstrup}}]{1989RScI...60.1026W}
{Wilkins}, S.~W., {Stevenson}, A.~W., {Nugent}, K.~A., {Chapman}, H., \&
  {Steenstrup}, S. 1989, Review of Scientific Instruments, 60, 1026,
  \dodoi{10.1063/1.1140312}

\bibitem[{{Willingale} \& {M{\'e}sz{\'a}ros}(2017)}]{2017SSRv..207...63W}
{Willingale}, R., \& {M{\'e}sz{\'a}ros}, P. 2017, \ssr, 207, 63,
  \dodoi{10.1007/s11214-017-0366-4}

\bibitem[{{Willingale} {et~al.}(2016){Willingale}, {Pearson}, {Martindale},
  {Feldman}, {Fairbend}, {Schyns}, {Petit}, {Osborne}, \&
  {O'Brien}}]{2016SPIE.9905E..1YW}
{Willingale}, R., {Pearson}, J.~F., {Martindale}, A., {et~al.} 2016, in Society
  of Photo-Optical Instrumentation Engineers (SPIE) Conference Series, Vol.
  9905, Space Telescopes and Instrumentation 2016: Ultraviolet to Gamma Ray,
  ed. J.-W.~A. {den Herder}, T.~{Takahashi}, \& M.~{Bautz}, 99051Y,
  \dodoi{10.1117/12.2232946}

\bibitem[{{Wu} {et~al.}(2022){Wu}, {Jia}, {Wang}, {Ling}, {Zhang}, {Zhang}, \&
  {Yuan}}]{2022PASP..134c5006W}
{Wu}, Q., {Jia}, Z., {Wang}, W., {et~al.} 2022, \pasp, 134, 035006,
  \dodoi{10.1088/1538-3873/ac5ac9}

\bibitem[{{Yuan} {et~al.}(2022){Yuan}, {Zhang}, {Chen}, \&
  {Ling}}]{2022arXiv220909763Y}
{Yuan}, W., {Zhang}, C., {Chen}, Y., \& {Ling}, Z. 2022, arXiv e-prints,
  arXiv:2209.09763.
\newblock \doarXiv{2209.09763}

\bibitem[{{Yuan} {et~al.}(2016){Yuan}, {Amati}, {Cannizzo}, {Cordier},
  {Gehrels}, {Ghirlanda}, {G{\"o}tz}, {Produit}, {Qiu}, {Sun}, {Tanvir}, {Wei},
  \& {Zhang}}]{2016grbu.book..237Y}
{Yuan}, W., {Amati}, L., {Cannizzo}, J.~K., {et~al.} 2016, in Gamma-Ray Bursts.
  Series: Space Sciences Series of ISSI, Vol.~61, 237--279,
  \dodoi{10.1007/978-94-024-1279-6_10}

\bibitem[{{Yuan} {et~al.}(2018){Yuan}, {Zhang}, {Ling}, {Zhao}, {Wang}, {Chen},
  {Lu}, {Zhang}, \& {Cui}}]{2018SPIE10699E..25Y}
{Yuan}, W., {Zhang}, C., {Ling}, Z., {et~al.} 2018, in Society of Photo-Optical
  Instrumentation Engineers (SPIE) Conference Series, Vol. 10699, Space
  Telescopes and Instrumentation 2018: Ultraviolet to Gamma Ray, ed. J.-W.~A.
  {den Herder}, S.~{Nikzad}, \& K.~{Nakazawa}, 1069925,
  \dodoi{10.1117/12.2313358}

\bibitem[{Zhang {et~al.}(2012)Zhang, Ling, \& Zhang}]{2012zhangSPIE}
Zhang, C., Ling, Z., \& Zhang, S.-N. 2012, in Space {Telescopes} and
  {Instrumentation} 2012: {Ultraviolet} to {Gamma} {Ray}, ed. T.~Takahashi,
  S.~S. Murray, \& J.-W. A.~d. Herder, Vol. 8443 (SPIE), 84433X,
  \dodoi{10.1117/12.925780}

\bibitem[{{Zhao} {et~al.}(2017){Zhao}, {Zhang}, {Yuan}, {Zhang}, {Willingale},
  \& {Ling}}]{2017ExA....43..267Z}
{Zhao}, D., {Zhang}, C., {Yuan}, W., {et~al.} 2017, Experimental Astronomy, 43,
  267, \dodoi{10.1007/s10686-017-9534-5}

\bibitem[{{Zhao} {et~al.}(2014){Zhao}, {Zhang}, {Yuan}, {Willingale}, {Ling},
  {Feng}, {Li}, {Ji}, {Wang}, \& {Zhang}}]{2014SPIE.9144E..4EZ}
{Zhao}, D., {Zhang}, C., {Yuan}, W., {et~al.} 2014, in Society of Photo-Optical
  Instrumentation Engineers (SPIE) Conference Series, Vol. 9144, Space
  Telescopes and Instrumentation 2014: Ultraviolet to Gamma Ray, ed.
  T.~{Takahashi}, J.-W.~A. {den Herder}, \& M.~{Bautz}, 91444E,
  \dodoi{10.1117/12.2055434}

\bibitem[{ZHAO {et~al.}(2019)ZHAO, WANG, ZHANG, CHEN, \&
  MA}]{Zhao2019Experiments}
ZHAO, Z.-j., WANG, Y.-s., ZHANG, L.-y., CHEN, C., \& MA, J. 2019, Optics and
  Precision Engineering, 27, 2330

\end{thebibliography}
\bibliographystyle{aasjournal}



\end{document}